\documentclass[printer]{gLMA2e}

\usepackage{rotating}
\usepackage{amsmath}
\usepackage{color}
\usepackage{xspace}

\newcommand{\R}[1]{\ensuremath{\mathbb R}^{\,#1}{}}
\newcommand{\C}[1]{\ensuremath{\mathbb C}^{\,#1}{}}
\newcommand{\Z}[1]{\ensuremath{\mathbb Z}_{\,#1}{}}

\newcommand{\HS}{\ensuremath{\mathcal H}{}}
\newcommand{\JZ}[1]{\ensuremath{J_{\nu_{#1}}^{(#1)}{}}}

\newcommand{\tr}{\operatorname {tr}}
\renewcommand{\Re}{\operatorname {Re}}
\renewcommand{\Im}{\operatorname {Im}}
\newcommand{\grad}{\operatorname{grad}}
\renewcommand{\mod}{\operatorname{mod}}

\newcommand{\comm}[2]{\ensuremath{[#1,#2]}}
\renewcommand{\vec}{\operatorname{vec}}

\newcommand{\bra}[1]{\ensuremath{\langle #1 |}{}}
\newcommand{\ket}[1]{\ensuremath{| #1 \rangle}{}}
\newcommand{\braket}[2]{\ensuremath{\langle #1 | #2 \rangle}{}}
\newcommand{\ketbra}[2]{\ensuremath{| #1 \rangle \langle #2 |}{}}
\newcommand{\expt}[1]{\ensuremath{\langle #1 \rangle}{}}

\newcommand{\unity}{\ensuremath{{\rm 1 \negthickspace l}{}}}

\newcommand{\Ad}[1]{\operatorname{Ad_{#1}}}
\newcommand{\adr}{\operatorname{ad}}
\newcommand{\Adr}{\operatorname{Ad}}

\newcommand{\Mat}{\operatorname{Mat}{}}
\newcommand{\diag}{\operatorname{diag}{}}
\newcommand{\Cd}{\ensuremath{C^{\dagger}}{}}
\newcommand{\Dd}{\ensuremath{D^{\dagger}}{}}
\newcommand{\Ed}{\ensuremath{E^{\dagger}}{}}
\newcommand{\Ud}{\ensuremath{U^{\dagger}}{}}

\newcommand{\GACk}{\ensuremath{\comm{U_k^{\phantom{\dagger}}AU_k^{\dagger}}{C_{\phantom{k}}^\dagger}}{}}
\newcommand{\GADk}{\ensuremath{\comm{U_k^{\phantom{\dagger}}AU_k^{\dagger}}{D_{\phantom{k}}^\dagger}}{}}
\newcommand{\TAC}{\ensuremath{f_C(U)}{}}
\newcommand{\TAD}{\ensuremath{f_D(U)}{}}
\newcommand{\TEE}{\ensuremath{f_E(U)}{}}

\newcommand{\TACc}{\ensuremath{f_C^*(U)}{}}
\newcommand{\TADc}{\ensuremath{f_D^*(U)}{}}
\newcommand{\TACck}{\ensuremath{f_C^*(U_k)}{}}
\newcommand{\TADck}{\ensuremath{f_D^*(U_k)}{}}
\newcommand{\GAC}{\ensuremath{\comm{UAU^{\dagger}}{C^\dagger}}{}}
\newcommand{\GAD}{\ensuremath{\comm{UAU^{\dagger}}{D^\dagger}}{}}
\newcommand{\GEE}{\ensuremath{\comm{UEU^{\dagger}}{E^\dagger}}{}}
\newcommand{\GEEk}{\ensuremath{\comm{U_k^{\phantom{\dagger}}EU_k^{\dagger}}{E^\dagger}}{}}

\newcommand{\pmi}{\ensuremath{\phantom{-}}}

\newcommand{\LAG}{{Lagrange}\xspace}

\newcommand{\wca}{\ensuremath{W(C,A)}\xspace}
\newcommand{\wda}{\ensuremath{W(D,A)}\xspace}

\newcommand{\wkca}{\ensuremath{W_{\mathbf K}(C,A)}\xspace}
\newcommand{\dwca}{\ensuremath{\partial{W(C,A)}}\xspace}

\newcommand{\rca}{\ensuremath{r(C,A)}\xspace}

\newcommand{\adj}[1]{\ensuremath{#1^{\dagger}{}}}

\newcommand{\iso}{\ensuremath{\overset{\rm iso}{=}}{}}
\newcommand{\corr}{\ensuremath{\hat =}{}}

\newcommand{\abs}[1]{\ensuremath{\vert #1 \vert{}}}
\newcommand{\Abs}[1]{\ensuremath{\left \vert #1 \right \vert{}}}
\newcommand{\norm}[1]{\ensuremath{\Vert #1 \Vert{}}}
\newcommand{\fnorm}[1]{\ensuremath{\Vert #1 \Vert{}}_2^{\phantom{2}}}
\newcommand{\fnormsq}[1]{\ensuremath{\Vert #1 \Vert{}}_2^2}

\newcommand{\maxover}[1]{\ensuremath{\underset{#1}\max}}
\newcommand{\minover}[1]{\ensuremath{\underset{#1}\min}}

\usepackage{german}
\begin{document}
\doi{10.1080/03081080xxxxxxxxxxxxx}
\issn{1563-5139}
\issnp{0308-1087} \jvol{00} \jnum{00} \jyear{2007} \jmonth{XXXX}

\markboth{T.~Schulte-Herbr{\"u}ggen et al.}%
{Significance of $C$-Numerical Ranges and Local $C$-Numerical Ranges in Quantum Control}

\title{The Significance of the $C$-Numerical Range and\\ the Local $C$-Numerical Range in Quantum Control and Quantum Information}

\author{Thomas Schulte-Herbr{\"u}ggen$^{*1}$\thanks{$^*$Corresponding author: %
\texttt{tosh@ch.tum.de}}, Gunther Dirr$^2$, Uwe Helmke$^2$, %
and Steffen J.~Glaser$^1$
%\thanks{\vspace{6pt}\newline\centerline{\tiny{
%ISSN 0308-1087 print/ ISSN 1563-5139 online
%\textcopyright 2007 Taylor \& Francis Ltd}}
%\newline\centerline{\tiny{ http://www.tandf.co.uk/journals}}\newline \centerline{\tiny{DOI:
%10.1080/03081080xxxxxxxxxxxxx}}},\\ %
$^1$ Dept. Chemistry, Technical University Munich, 85747 Garching, Germany\\
$^2$ Dept. Mathematics, University of W{\"u}rzburg, 97074 W{\"u}rzburg, Germany }  
\received{dated \today}

\maketitle

\begin{abstract}
This paper shows how $C$-numerical-range related new strucures may arise
from practical problems in quantum control---and vice versa, how an understanding of
these structures helps to tackle hot topics in quantum information.

We start out with an overview on the role of $C$-numerical ranges in current research
problems in quantum theory: the quantum mechanical task of maximising the projection
of a point on the unitary orbit of an initial state onto a target state $C$
relates to the $C$-numerical radius of $A$ via maximising the trace function
$|\tr \{C^\dagger UAU^\dagger\}|$. In quantum control of $n$ qubits one may be
interested (i) in having $U\in SU(2^n)$ for the entire dynamics, or (ii) in restricting
the dynamics to {\em local} operations on each qubit, i.e. to the $n$-fold tensor
product $SU(2)\otimes SU(2)\otimes \cdots\otimes SU(2)$.
Interestingly, the latter then leads to a novel entity, the {\em local} $C$-numerical range
$W_{\rm loc}(C,A)$, whose intricate geometry is neither star-shaped nor simply connected
in contrast to the conventional $C$-numerical range. This is shown in the
accompanying paper 
on {\em Relative $C$"~Numerical Ranges for Application in Quantum Control and Quantum Information} \cite{LAMA_WKCA}. 
We present novel applications
of the $C$-numerical range in quantum control assisted by gradient flows on the local
unitary group: (1) they serve as powerful tools for deciding whether a quantum interaction
can be inverted in time (in a sense generalising Hahn's famous spin echo); (2) they allow for
optimising witnesses of quantum entanglement.
We conclude by relating the relative $C$"~numerical range to problems of constrained quantum optimisation,
for which we also give \LAG-type gradient flow algorithms.
\medskip
\end{abstract}

%%%%%%%%
{\quote\small%
{\em We are currently in the midst of a second quantum revolution. The\\
        first one gave us new rules that govern physical reality. The second\\
        one will take these rules and\, use them to develop new technologies.\\
        \phantom{X}\hspace{53mm}\small{\rm Dowling and Milburn, 2003 \cite{DowMil03}}}
}\\[2mm]
%%%%%%%%
It was the main goal of the talk entailing this paper to
entice the numerical range community to showing
further interest in problems of optimisation and control of quantum systems.
By illustrating how important properties of 
$C$"~numerical ranges relate to
reachability and optimisation in quantum dynamics, we wish to foster cross-fertilisation 
leading---in turn---also to new discoveries in numerical ranges.
These may go beyond or follow earlier work on applying 
higher-rank numerical ranges \cite{LiTsi91} to quantum error correction \cite{Choi++06},
or
numerical ranges of derivations to anti-symmetric
quantum states \cite{LiProv93}, on optimising coherence transfer in 
quantum systems \cite{Science98, TOSH-Diss, NMRJOGO}, which entailed special
interest in the $C$"~numerical range of nilpotent matrices relevant in spectroscopy \cite{DHK06, LiWoerd06}. 
As contribution to our end, we present some new results on what we introduce 
as {\em relative $C$"~numerical range} \cite{LAMA_WKCA},
we relate some of our recent results in quantum control to numerical ranges, and we express
corner stones of quantum control in the setting of numerical ranges. A reader interested in the
quantum aspects may appreciate the paper being organised such as to pursue these issues 
in reverse order, whereas the one driven by impatient curiosity may prefer to jump into
Section \ref{sec:alg_WCA} or Chapters \ref{chp:loc_WCA} and \ref{chp:const_WCA} right away.
\vspace{-2mm}

\section{Overview on $C$"~Numerical Ranges in Quantum Control}
Controlling quantum systems offers a great potential
for performing computational tasks or for simulating the behaviour of other quantum systems
\cite{Fey82,Fey96}. This is because the complexity of many problems \cite{Pap95} reduces upon going
from classical to quantum hardware. It roots in Feynman's observation \cite{Fey82}
that the resources required for simulating
a quantum system on a classical computer increase exponentially with the system size. In turn,
he concluded that using quantum hardware might therefore {\em exponentially decrease}
the complexity of certain classical computation problems. Coherent superpositions of quantum states
used as so-called \/`qubits\/' can be viewed as a particularly powerful resource of quantum
parallelism unparalleled by any classical system.
Important applications are meanwhile known in quantum computation,
quantum search and quantum simulation:
most prominently, there is the exponential speed-up by Shor's quantum algorithm
of prime factorisation \cite{Shor94short,Shor97}, which
relates to the general class of quantum algorithms \cite{Jozsa88,Mosca88}
solving hidden subgroup problems in an efficient way \cite{EHK04}.

However, for exploiting the power of quantum systems, one has to steer them by classical
controls such as voltage gates, radio-frequency pulses, or laser beams. 
It is highly desirable to do so in an optimal way, because the shapes
of these controls critically determine the performance of the quantum system in terms
of overlap of its actual final states with the desired target states.
Here, the  aim is to show how quantum optimal control relates to finding
points on the unitary orbit of the initial quantum state $A$ (in its density matrix representation) 
maximally projecting onto the desired target state $C$
which is equivalent to maximising the target function 
\begin{equation}
F(U)=|\tr\{UAU^\dagger C^\dagger\}|
\end{equation}
over all unitaries to yield the {\em $C$"~numerical radius of $A$}.
This is the scope in systems that are 
fully controllable in the sense that every propagator
$U\in SU(2^n)$ can be realised on the physical system in question.

Yet, often only local unitary operations $K\in SU(2)\otimes SU(2)\otimes\cdots\otimes SU(2)$
are actually available. Then the corresponding optimisation problems are confined to a subset
of the conventional $C$"~numerical range: this is the motivation to introduce the {\em local $C$"~numerical range}
$W_{\rm loc}(C,A)$ as a special case of the {\em relative $C$"~numerical range} \wkca,
whose intricate geometry is analysed in more detail in the accompanying paper \cite{LAMA_WKCA}.

In view of practical applications in quantum control, we finally give an outlook
on contrained optimisation problems, i.e., those in which extremal points in the
$C$"~numerical range are searched subject to fulfilling contraints such as
keeping $UAU^\dagger$ orthogonal to an undesired state $D$ or leaving a neutral
state $E$ invariant.

Since in general,
%algebraic optimisation
a precise characterisation of solutions in algebraic terms is often beyond
reach, we resort to numerical methods based on gradient flows on the unitary
group.

\subsection{Quantum Dynamics: Notations and Relation to $C$-Numerical Range}\label{sec:qd_notation}
As usual in quantum mechanics, one may choose to represent a state of a pure quantum system
by a state vector in Hilbert space, $\ket\psi\in\mathcal H$. Its norm induced by the scalar
product can be set to $\braket{\psi}{\psi}=1$, as will be assumed henceforth. 
The operators associated to quantum
mechanical observables such as the Hamiltonian $H$ are selfadjoint,
%and bounded (??? Im Allgemeinen ist $H$ nur abgeschlossen aber NICHT
% beschr"ankt. ???)
so $H=H^\dagger$. %\in \mathcal B(\mathcal H)$.
Then the Hamiltonian dynamics is governed by Schr{\"o}dinger's equation of
motion 
\begin{equation}
        \ket{\dot\psi(t)} = -i H   \;\ket{\psi(t)} \quad\text{solved by}\quad
        \ket{\psi(t)} = e^{-i tH}  \;\ket{\psi(0)} = {U(t)} \;\ket{\psi(0)} \;.
\end{equation}
Invoking Stone-von Neumann's theorem, 
the solution involves a time evolution
by an element $U(t)$ of the strongly continuous one-parameter group 
$\mathcal U:=\{e^{-itH}\,|\,t\in\R{}\}$
generated by the Hamiltonian $H$, see e.g. \cite{Dav80}.

Non-pure quantum states comprise the settings of classically mixed states as well as reduced
representations of quantum systems allowing for the description of open dissipative systems.
In these cases, one
may choose to represent the state by a positive-semidefinite trace-class operator, the {\em density operator}
$\rho \in \mathcal B_1(\mathcal H)$ with $\rho \geq 0$ being normalised
to $\tr \rho =1 $. With the trace-class operators $\mathcal B_1(\mathcal H)$ forming a two-sided 
ideal in the bounded ones $\mathcal B(\mathcal H)$,
its dynamics follows Liouville-von Neumann's equation
        \begin{equation}
        \dot \rho(t) = -i \comm{H}{\rho(t)} \quad\text{solved by}\quad
        \rho(t) = {U}(t) \rho(0) {U(t)^\dagger}\quad,\quad
        %\in \mathcal U\big(\rho(0)\big)
        U(t) = e^{-itH}
        \end{equation}
and thus \/`dwells'\/ on the {unitary orbit} of the initial state
$\mathcal O_{u}\big(\rho(0)\big)$.  

Next, consider the expectation value of observables $B = B^\dagger \in \mathcal B(\mathcal H)$
in either setting. For pure quantum states it takes the form of the scalar product $\braket \cdot \cdot$
        \begin{equation}
        \expt B(t) := \braket{\psi(t) B}{\psi(t)}
        %\quad\in\quad    {W(B)}:= \{u^\dagger B u\,\big| \norm{u}=1\}
        \end{equation}
while for non-pure states (using the Hilbert-Schmidt analogue $\tr\{\cdot^\dagger \cdot\}$)
it reads
        \begin{equation}
        \expt{\expt B}(t) = \tr\big(B^\dagger \rho(t)\big) = 
        \tr \big(B^\dagger\; U(t)\rho(0) U(t)^\dagger\big)\quad.
        \end{equation}
Clearly, in pure states the expectation value is an element of the field of values 
$\expt B(t) \in W(B):=\{\braket{u}{Bu}\,\big| \norm{u}=1\}$, whereas in non-pure states
it is an element of the $C$-numerical range $\expt{\expt B}(t) \in W(B,\rho(0))$ then
taking the form of a real line segment.
The latter is of particular significance, e.g., in ensemble spectroscopy, where it is
customary to collect all signal-relevant components of the
selfadjoint operator $B$ in a new operator $C$ that need no longer be
Hermitian,
and likewise the pertinent terms of $\rho(0)$ in a general complex operator
$A$.  
Thus moving from the selfadjoint operators $B,\rho$ to arbitrary, not
necessarily Hermitian (yet bounded) operators $A,C$, the analogue to the
ensemble expectation value then becomes a general
element of the $C$-numerical range $W(C,A) := \{\tr (C^\dagger\; U A U^\dagger)| U\in\mathcal U(\mathcal H)\}$,
where here and henceforth $A,C\in\Mat_n(\C{})$ are taken to be finite-dimensional
in view of spin and pseudo-spin systems.

The key features of \wca \cite{GS-77,Li94} 
may thus be exploited for quantum optimisation and control. They comprise:
(i) the $C$"~spectrum of $A$ is a subset of \wca;
(ii) \wca is always compact, connected and star-shaped \cite{TSING-96}
        with respect to the centre $\tr\{A\}\tr\{C^\dagger\}/N$;
        it is convex if (but not only if) $C$ is normal with collinear
        eigenvalues in the complex plane, or if there is a $\mu\in\C{}$ so that $(C-\mu\unity)$
        has rank 1;
        it is a circular disc in the complex plane \cite{Li91} if
        there is a $\mu\in\C{}$ so that $(C-\mu\unity)$ is unitarily similar
        to block-shift form;
(iii) the corners of the boundary \dwca at which no tangent exists are always elements of the $C$"~spectrum
        of $A$;
(iv) for normal $C$ with collinear eigenvalues in the complex plane (as well as in some degenerate
        cases \cite{Li94}) \wca is the closed convex hull of the $C$"~spectrum of $A$ thus 
        forming a convex polygonal disc in the complex plane.
%%%%

\subsection{Geometry of Optimisation within $C$"~Numerical Ranges}
In the context of $C$-numerical ranges, there are geometric optimisation tasks immediately
related to problems of quantum control, e.g., finding
points on the unitary orbit of (the initial state) $A$ that
\begin{enumerate}
\item[(1)] show a minimum Euclidean distance to (the target state) $C$ corresponds to
        the maximum real part of the $C$-numerical range by 
        \begin{equation}\label{eqn:distance}
        {|| C - UAU^\dagger ||}_2^2 = {||A||}_2^2 + {||C||}_2^2 - 2 \Re \tr \{C^\dagger\; U A U^\dagger\}\quad,  
        \end{equation}
        while those that
\item[(2)] enclose a minimal angle (mod $\pi$) to $C$ relate to the
        $C$-numerical radius $\rca := \maxover U |\tr\{C^\dagger UAU^\dagger\}|$ by
        \begin{equation}\label{eqn:angle}
        \cos^2_{A,C}\,(U) = \frac{|\tr\{C^\dagger UAU^\dagger\}|^2}{{\|A\|}_2^2\cdot{\|C\|}_2^2}\quad.  
        %\cos^2_{A,C}\,(U) = {|\tr\{C^\dagger UAU^\dagger\}|^2}/\big({{\|A\|}_2^2\cdot{\|C\|}_2^2}\big)\quad.  
        \end{equation}
\end{enumerate}
Clearly, the mathematical limits to unitary transfer from $A$ onto $C$ are physically
meaningful only if all the transformations in the entire unitary group can be realised
in the given experimental setting. This is what we will analyse in the next section.---
In a fully controllable system, the $C$-numerical radius coincides with the maximal
transfer of relevant components collected in $A$ onto %those of 
the target $C$. In coherent ensemble spectroscopy, this is identical
to the maximal spectroscopic signal amplitudes obtainable
in the absence of relaxation \cite{Science98,TOSH-Diss}.

\subsection{Controllability of Quantum Systems}
The standard {\em bilinear control system} with state $X(t)$,
drift $A$, controls $B_j$, and control amplitudes $u_j\in\R{}$
reads 
\begin{equation}
        \dot X(t) = \big(A + \sum_{j=1}^m u_j(t) B_j\big) \; X(t) \quad,
\end{equation}
while $X(t) \in \mathcal{GL}_N(\C{})$ and $A,B_j\in \Mat_N(\C{})$.
Thus Hamiltonian quantum dynamics following Schr{\"o}dinger's equation
\begin{eqnarray}
        \ket{\dot\psi(t)} &=& -i\big(H_d + \sum_{j=1}^m u_j(t) H_j\big) \;\ket{\psi(t)}\\
        \label{eqn:bilinear_contr}
        {\dot U(t)} &=& -i\big(H_d + \sum_{j=1}^m u_j(t) H_j\big) \;{U(t)} \quad,
\end{eqnarray}
can be cast into the above setting. Here $H_d$ is the drift term,
$H_j$ are the control Hamiltonians with $u_j(t)$ as control amplitudes.
%In systems of 
For $n$ qubits, $\ket{\psi}\in \C{2^n}$,
$U\in SU(2^n)$, and $i\,H_\nu\in\mathfrak{su}(2^n)$.% for $\nu\in\{d,j\}$.

\begin{definition}[(Full Controllability)]\\
A system is {\em fully controllable}, if to every initial state $A$ the entire unitary
orbit $\mathcal O_u(A)$ can be reached. 
\end{definition}
In the special case of Hermitian operators
this means any final state $X(t)=:C$ can be reached from any initial state $X(0)=:A$ 
as long as the operators $A$ and $C$ share the same spectrum of eigenvalues. % (in finite time).

\begin{corollary}{\rm \cite{SJ72JS}}\label{cor:controllability1}\\
The bilinear system of Eqn.~\ref{eqn:bilinear_contr} is {fully controllable} if and only if
drift and controls are a generating set of $\mathfrak{su}(2^n)$ by
way of the commutator, in the sense
$\langle{H_d, H_j} \,|\,j=1,2,\dots,m\rangle_{\rm Lie} = {\mathfrak{su}(2^n)}$.
\end{corollary}

\begin{example}[($n$ weakly coupled spin-$\tfrac{1}{2}$ qubits):]\\
%%%%%%%%%%%%%%%%
Let $\sigma_x = \left(\begin{smallmatrix} \,0 &\,1\, \\ \,1 &\,0\, \end{smallmatrix}\right)$,
$\sigma_y = \left(\begin{smallmatrix} 0 &-i \\ i &\phantom{-}0 \end{smallmatrix}\right)$,
$\sigma_z = \left(\begin{smallmatrix} 1 &\phantom{-}0 \\ 0 &-1\end{smallmatrix}\right)$
be the Pauli matrices.
In %a system of
$n$ spins"~$\tfrac{1}{2}$,
a $\sigma_{kx}$ for spin $k$ is tacitly embedded as
$\unity\otimes\cdots\unity\otimes\sigma_{x}\otimes\unity\otimes\cdots\unity$
where $\sigma_{x}$ is at position $k$.
The same holds for $\sigma_{ky}$,  $\sigma_{kz}$, and in the weak coupling
terms $\sigma_{kz}\sigma_{\ell z}$ with $1\leq k<\ell\leq n$.
%%%%%%%%%%%%%%%

\begin{theorem}[(Controllability of Coupled Qubits)]{\rm \cite{TOSH-Diss}}\\
A system of $n$ %(non-degenerate)
qubits is {fully controllable},
if e.g. the control Hamiltonians {$H_j$} comprise the Pauli matrices 
$\{{\sigma_{kx}, \sigma_{ky}}\,|\, k=1,2,\dots n\}$
on every single qubit selectively and the drift Hamiltonian {$H_d$}
encompasses the Ising pair interactions
$\{{J_{k\ell}} \; {(\sigma_{kz}\sigma_{\ell z})/2}\,|\, k<\ell=2,\dots n\}$, 
where the coupling topology of $J_{k\ell}\neq 0$
may take the form of {any connected graph}.
\end{theorem}
This theorem has meanwhile been generalised to other types of couplings
\cite{GA02,AlbAll02}.
\end{example}

\begin{example} [(Quantum Gates)]: \\
In quantum computing, the logical gate operations have a unitary representation.
Therefore, implementing a unitary gate by a sequence of evolutions under drift and control
terms of the respective hardware (i.e., the physical quantum system) can be seen as
the {\em quantum compilation} task: it translates the unitary gates into the machine code
of physically accessible controls.
\end{example}

\begin{corollary}
The following are equivalent:
\begin{enumerate}
\item[(1)] in a quantum system of $n$ coupled spins-$\tfrac{1}{2}$, the drift $H_d$ and the controls $H_j$
        form a generating set of $\mathfrak{su}(2^n)$;
\item[(2)] every unitary transformation $U\in SU(2^n)$ can be realised by that system;
\item[(3)] there is a set of {universal quantum gates} for the quantum system;
\item[(4)] the quantum system is fully controllable;
\item[(5)] the reachability set to the generalised 
        expectation value 
        $\expt{\expt{C}}(t)=\tr\{C^\dagger A(t)\}$
       coincides with the {$C$-numerical range $W(C,A)$} for all
       $A,C \in \Mat_{2^n}(\C{})$.
\end{enumerate}
\end{corollary}
\begin{proof}:
(1) $\Leftrightarrow$ (2) follows by
using the surjectivity of the exponential map in compact connected groups \cite{SJ72JS};
%the rest is trivial:
(2) $\Leftrightarrow$ (3) express the same fact in the terminology of group theory (2) and
quantum computing (3); (4) $\Leftrightarrow$ (1) follows by Corollary~\ref{cor:controllability1}; 
(4) $\Leftrightarrow$ (5) by definition of full controllability via reachability of 
the entire unitary orbit $\mathcal O_u(A)$.
\end{proof}

\subsection{Tasks in Optimal Quantum Control}
`\/Optimise a scalar quality function subject to the equation of motion
governing the dynamics of the system to be steered\/'--this is the formal setting
of many an engineering problem both in classical and quantum systems.

Extending the notions of quantum dynamics introduced above in Sec.~\ref{sec:qd_notation}, 
there are two principle types of scenarios, (i) closed Hamiltonian systems evolving
without relaxation and (ii) systems open to dissipation.
Define $U(t):=e^{-it H}$ and the unitary conjugation map $\Adr_U(\cdot):= U(\cdot)U^{-1}$ as well as
the commutation operator $\adr_H(\cdot):= [H,\cdot]$ to obtain the following equations of motion:\\[3mm]
%%%%%%%%%%%%%%%
{\em (i) closed Hamiltonian systems}\vspace{-3mm}
\begin{alignat}{6}\label{eqn:psidot}
&\text{1. pure state}& \dot{\ket\psi} &= -i H\;
\ket\psi\hspace{1cm}&\ket\psi(t)&\in\mathcal H \iso \C{}^N&\hspace{.5cm}\\\label{eqn:Udot}
&\text{2. gate}& \dot{U} &= -i H\; U&U(t)&\in\mathcal U(\mathcal H)& \\\label{eqn:rhodot}
&\text{3. non-pure state}& \dot{\rho} &= -i \adr_H\;(\rho) &\rho(t)&\in\mathcal B_1(\mathcal H)& \\\label{eqn:AdUdot}
%&\text{4. projective gate\quad}& \dot{\chi} &= -i \adr_H\;\circ\;\chi& \chi(t) &= \in\mathcal U\big(\mathcal B_1(\mathcal H)\big)&
&\text{4. projective gate\quad}& \dot{\Adr}_U &= -i \adr_H\;\circ\;\Adr_U\,,\quad&\Adr_U(t)&\in\mathcal U\big(\mathcal B_1(\mathcal H)\big)& 
\end{alignat}
%%%
%%%
{\em (ii) open dissipative systems}\vspace{-3mm}
\begin{alignat}{6}\label{eqn:rhodot2}
&\text{3'. non-pure state}& \dot{\rho} &= -(i \adr_H
\,+\,\Gamma)\;(\rho)\hspace{0.9cm}&\rho(t)&\in\mathcal B_1(\mathcal H)&
\\\label{eqn:FAdUdot}
%&\text{4'. {contractive} map\quad}& \dot{\chi} &= -(i \adr_H \,+\,\Gamma)\;\circ\; \chi\;& \chi(0)&= \Adr_{U(0)}&
%\in\mathcal{GL}\big(\mathcal B_1(\mathcal H)\big)&,
&\text{4'. {contractive} map\quad}& \dot \chi &= -(i \adr_H \,+\,\Gamma)\;\circ\; \chi,\;& \chi(t)&\in\mathcal{GL}\big(\mathcal B_1(\mathcal H)\big)&,
%&\text{4'. {contractive} image\quad}& \dot{F}_{\Adr_{U}} &= -(i \adr_H \,+\,\Gamma)\;\circ\; F_{\Adr_U},\;& F_{\Adr_U}&\in\mathcal{GL}\big(\mathcal B_1(\mathcal H)\big)&,
\end{alignat}
where $\mathcal H$ is a finite dimensional Hilbert space while
$\mathcal U(\mathcal H)$ and $\mathcal B_1(\mathcal H)$ denote
the respective unitary group as well as the trace-class operators over
$\mathcal H$, and $\mathcal{GL}\big(\mathcal B_1(\mathcal H)\big)$ is the 
general linear group over $\mathcal B_1(\mathcal H)$. 
%%%%%%%%%%%%%%%
Note that Eqn.~\ref{eqn:Udot} is the operator equation to Eqn.~\ref{eqn:psidot} referring to the
unitary map of the entire basis.
Likewise, for the unitary conjugation map, Eqn.~\ref{eqn:AdUdot} is the operator equation to Eqn.~\ref{eqn:rhodot}.
If the density operator $\rho$ is viewed as a vector in Liouville space, e.g. by way
of the $\vec$ representation \cite{HJ2}, 
then the map $\Adr_U$ is an element of the {\em projective special unitary group}
\begin{equation}
\Adr_U \in PSU(N) \iso \frac{U(N)}{U(1)}\iso\frac{SU(N)}{\Z N}\quad,
\end{equation}
where $\Z N$ denotes the centre of $SU(N)$.
At the expense of being highly reducible, one may choose the embedded representation 
$(U^* \otimes U)\in SU(N^2)$ for the convenience of having $(U^* \otimes U) \vec(\rho)\, \corr\, U\rho\, U^\dagger$.

\noindent
When including dissipation via the positive semidefinite
relaxation operator $\Gamma\geq 0$, the operator equation (Eqn.~\ref{eqn:FAdUdot})
to the Master equation (Eqn.~\ref{eqn:rhodot2}) describes the contractive
quantum map $\chi(t)$ generalising the unitary conjugation map $\Adr_U$ for open dissipative systems. 

The scenarios of Eqns.~\ref{eqn:Udot}, \ref{eqn:AdUdot}, \ref{eqn:FAdUdot}
 occur in the following typical optimisation problems of quantum control for (tracking over) fixed final times $T$:
\begin{itemize}
        \item[A] {\em Maximise Experimental Sensitivity in Coherent Spectroscopy
                  by Finding the $C$-Numerical Radius of $A$} \cite{Science98,TOSH-Diss}:\\
                maximise transfer amplitude
                        $f(U) := \abs{\tr\{C^\dagger  A(T)\}} = \abs{\tr\{C^\dagger U A U ^\dagger\}}$\\[1mm]
                        subject to equation of motion
                        $ \dot U(t) = -i {H}{U(t)}$\\[1mm]
        \item[B] {\em Realise Unitary Module $U_{\rm G}$ in Minimal Time:}\cite{Khaneja02,PRA05}\\
                maximise fidelity $f(\Adr_U(T)):= \Re \tr \{\Adr_{U_G}^\dagger \Adr_U(T)\}$\\
                subject to equation of motion
                $ \dot{\Adr_U}(t) = -i \adr_H\,\circ\,\Adr_U(t)$\\[1mm]
        \item[C] {\em Approximate Unitary Module $U_G$ with Minimal Relaxative Loss:}\cite{PRL_decoh,PRL_decoh2}\\
                maximise fidelity $f(\chi(T)):= \Re \tr \{\Adr_{U_G}^\dagger \chi(T)\}$\\[1mm]
                subject to Master equation of motion
                $\dot \chi(t) = -(i \adr_H + \Gamma )\,\circ\,\chi(t)$\;
\end{itemize}

Here we focus on problem A which determines the limit to unitary transfer
on a general abstract level. It relates to the $C$-numerical radius in the
fully controllable case and to the relative $C$-numerical radius
\cite{LAMA_WKCA} in the non-controllable case. In view of experimental
implementation, in a second step, the family of critical unitary operators 
\begin{equation}
\mathcal U_0 :=\{U\;\big |\; |\tr(C^\dagger\,UAU^\dagger)| = r_C(A)\}
\end{equation}
may be realised or approximated in concrete experimental settings either in
the fastest way (task B) or with least amount of relaxative loss (task C).

\subsection{Gradient Flows Determining the $C$-Numerical Range and Radius}\label{sec:alg_WCA}
In this section, we refer to numerical algorithms based on gradient-flows 
for obtaining the $C$-numerical radius, which means solving task A in
the fully controllable case.

If $A,C\in \Mat_n(\C{})$ are Hermitian, the $C$-numerical range of
$A$ is a real line segment. Its maximum results from sorting
the eigenvalues of $A,C$ by magnitude in same order as has been
shown by von Neumann in 1937 \cite{NEUM-37} and in view of NMR spectroscopy
by S{\o}rensen \cite{OLE-89}.
For the special case of real symmetric matrices, a gradient flow
on the group of special orthogonal matrices $SO(N)$ was presented
in the pioneering work of Brockett \cite{Bro88+91}, a thorough analysis of which
with convergence-ensuring step sizes for the numerical discretisation schemes
can be found in the monography of Helmke and Moore \cite{Helmke94}.
While in the Hermitian case, the gradient flows generically converge to {\em global}
extrema, an analogous result for the more general case,
where $A,C\in \Mat_n(\C{})$ may be arbitrary complex matrices, 
is still missing, since it appears much more involved. 
However, the gradient flows may be generalised as %has been 
shown in \cite{Science98,NMRJOGO},
and in all the cases we have been addressing over the years, the maxima found numerically 
have been on the boundary \dwca as conjectured in \cite{Science98}.

With $U\,:=\,e^{-i\,t\,H}$ and $f(U)\,:=\,\tr\{\adj C\,UA\adj U\}$, define the target functions
\begin{equation}
F_1(U)\,:=\,\Re f(U)\quad\text{and}\quad F_2(U)\,:=\,\Abs{f(U)}^2\;.
\end{equation}
For $\nu=1,2$ one finds the 
Fr{\'e}chet derivatives of $F_\nu$ at $U\in U(N)$ as the linear maps on the tangent space
$T_UU(N)$  comprising elements of the form $(iHU)$
%%%%%%%%%
\begin{eqnarray}\label{DfU}
D F_\nu(iHU)\,\,&=\,\,\tr\{&G^{(\nu)}\,iH\}\\
&& G^{(1)}=\tfrac{1}{2}\,\{\comm{UAU^\dagger}{C^\dagger}\,\,-\,\,\adj{\comm{UAU^\dagger}{C^\dagger}} \}\\
&& G^{(2)}=\comm{UAU^\dagger}{C^\dagger}\,f(U)^*\,\,-\,\,\adj{\comm{UAU^\dagger}{C^\dagger}}\,f(U)\quad\\
\label{eq:grad}
\text{\rm so}\quad \grad F_\nu(U)&=&-G^{(\nu)}\,U
\end{eqnarray}
is the respective gradient with $G^{(\nu)}$ skew-Hermitian.
By compactness of $U(N)$ the solutions of Eqn.~\ref{eq:gradII} exist for all
$t\in\R{}$ and converge to the set of critical points since 
$\grad F_\nu$ is a real analytic gradient vector field \cite{Helmke94}.

Clearly, $D F_\nu(iHU) = 0 $  in any direction $H$ implies $G^{(\nu)}=0$.
One may integrate the respective
differential equation
\begin{equation}
\label{eq:gradII}
\dot U = \grad F_\nu(U)
\end{equation}
to arrive at the recursive scheme
\begin{eqnarray}
\phantom{\sum\limits_M^N}U_{k+1}^{(\nu)}\,&=&\,e^{-\alpha_k^{(\nu)} G_k^{(\nu)}}\,U_k^{(\nu)}\qquad\text{till}%
                \qquad\big\|G^{(\nu)}_{k+1}\big\|_2 \to 0\quad.
\end{eqnarray}
As will be shown next, this gradient flow can readily be adapted to visualise the actual shape of \wca.

%%%%%%%%%%%%%%%% THESIS %%%%%%%%%%%%%%%%
%%%%%%%%%%%%%%%%%%%%%%%%%%%%%%%%%%%%%%%%
%%%%%%%%%%%%%%%%%%%%%%%%%%%%%%%%%%%%%%%%
\begin{algorithm*}{\rm (Determining the Boundary \dwca \cite{Science98,TOSH-Diss}}):\\
The star-shapedness of the $C$-numerical range of $A$ 
with respect to the centre $\tr{A}\cdot\tr{C^\dagger}/N$\cite{TSING-96}
is central for the following straightforward gradient algorithm to
determine the shape of \wca by (best approximations to) its boundary \dwca: 
\begin{enumerate}
\item[(1)] shift the reference frame to the centre of the star: $A\mapsto A-\tfrac{\tr\{A\}}{N}\,\unity$;
\item[(2)] modify the above gradient algorithm to drive into
        the intersection of \dwca with the positive real axis
        by the \LAG approach described in the next paragraph;
\item[(3)] rotate \wca stepwise in the complex plane by way of
         multiplying say matrix $A$ with a phase factor 
        $e^{i\,\ell 2\,\pi/m}$;
\item[(4)] repeat steps (2) and (3) for $\ell=1,2, \dots, m$;
\item[(5)] retransform the results in (2) into the original reference frame:
        the intersection points then give an approximation of the circumference \dwca.
\end{enumerate}
\end{algorithm*}

%%%%%%%%%%%%%%%%%%%%%%%%%%%%%%%%%%%%%%%%
%%%%%%%%%%%%%%%%%%%%%%%%%%%%%%%%%%%%%%%%
\noindent
Step (2) comprises a constrained optimisation implemented by \LAG multipliers as described next.\\[2mm]
\noindent
{\em Lagrange Approach to Tracing \dwca} \cite{TOSH-Diss}:\\[1mm]
Let again $f(U)\,:=\,\tr\{\adj C\,UA\adj U\}$ with $U \in U(N)$,
where we assume without loss of generality the reference frame has been
chosen such that $A$ is traceless so the star centre coincides with the origin.
In order to find the intersection of \dwca with the positive real axis, 
one has to maximise $F_1(U):=\Re f(U)$ while keeping $F_2(U):=\Im f(U)$ zero.
To this end, we introduced the Lagrange function
\begin{equation}
L(U) := F_1(U) - \lambda \big(F_2(U)\big)^2\quad,
\end{equation}
with $\lambda$ as multiplier.
Its Fr{\'e}chet derivative has the components
\begin{eqnarray}
D\{F_1(U)\}\,(iHU)&=&\tfrac{1}{2}\, \tr\{(\comm{UA\Ud}{C^\dagger}-\comm{UA\Ud}{C^\dagger}^{\dagger})\, iH\}\\
D\{F_2(U)\}^2\,(iHU)&=&-i\,\big(F_2(U)\big)\, \tr\{\comm{UA\Ud}{C^\dagger}+\comm{UA\Ud}{C^\dagger}^{\dagger})\, iH\}\qquad
\end{eqnarray}
so that one obtains the adapted recursive scheme \cite{TOSH-Diss}
\begin{equation}
U_{k+1} = \exp\{-\alpha\,\big(\GACk_S\, +2i\,\lambda\,\big(F_2(U_k)\big) \GACk_H\big)\} \, U_k\quad,
\end{equation}
where for short, $[\cdot,\cdot]_S$ and $[\cdot,\cdot]_H$ denote the skew-Hermitian and the Hermitian part
of the commutator, respectively.
%%%%%%%%%%%%%%%%%%%%%%%%%%%%%%%%%%%%%%%%
%%%%%%%%%%%%%%%%%%%%%%%%%%%%%%%%%%%%%%%%

\begin{figure}[Ht!]
\begin{center}
\includegraphics[width=5cm]{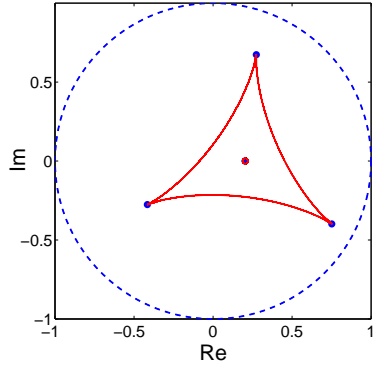}
\end{center}
\caption{\label{fig:WCA3} (Colour online) The $C$"~numerical range for $A,C$ of Example~\ref{ex:WCA3} 
        using the \LAG-type gradient-flow algorithm \cite{TOSH-Diss} described in the text. 
        The dots give the $C$-spectrum of $A$, where the interior ones coincide with the star centre.}
\end{figure}

\begin{example} \cite{Science98}\label{ex:WCA3}
Define the following pair of $3\times 3$ matrices
\begin{equation}
\begin{split}
A &:=\diag\big((0.7385 + 0.2400 i), (0.0353 - 0.1660 i), (0.4509 + 0.4060 i)\big)\quad\text{and}\quad\\
C &:=\diag\big((0.7189),(-0.1106 + 0.4789 i),(-0.1106 - 04789 i)\big)\quad.
\end{split}
\end{equation}
Fig.~\ref{fig:WCA3} shows the $C$"~numerical range $\wca := \tr\{UA\Ud\Cd\}$, where the concave
triangle is a particular challenge to the algorithm, since it requires 500 points to determine
the circumference reaching into the vertices. The \LAG parameter is dynamically increased 
with the iterations $k$ \cite{TOSH-Diss}. 
%: it follows the incomplete $\Gamma$-function (??? Was
%ist die 'incomplete $\Gamma$-function'???) smoothly from
%$\lambda=20$ to $\lambda=5000$
%with a turning point after $k=200$ iterations.
Note that the vertex points derived from the $C$"~spectrum of $A$ are perfectly reached.
Meanwhile, the shape has been quantitatively corroborated by global 
optimisation methods---as has also been shown during the {\sc wonra} in a collaboration with Prof.~Tibken's
group.
\end{example}
%%%%%%%%%%%%%%%%%%%%%%%%%%%%%%%%%%%%%%%%
%%%%%%%%%%%%%%%%%%%%%%%%%%%%%%%%%%%%%%%%

%%%%%%%%%%%%%%%%%%%
%%%%%%%%%%%%%%%%%%%

\section{Local $C$-Numerical Ranges}\label{chp:loc_WCA}
In view of applications in quantum control, it is customary to term the $n$"~fold
tensor product $SU(2)\otimes SU(2) \otimes \cdots \otimes SU(2)$ 
as the group of {\em local unitary operations} $SU_{\rm loc}(2^n)$. It is a subgroup to the full dynamic group
$SU(2^n)$. Consequently, to a given initial state $A$, the reachability set under
local controls amounts to the {\em local unitary orbit}
$\mathcal O_{\rm loc}(A):=\{KAK^\dagger | K \in  SU_{\rm loc}(2^n)\}$.

\begin{definition}
As in the companion paper \cite{LAMA_WKCA}, we define as {\em local
  $C$-numerical range} of $A$ the subset
\begin{equation}
 W_{\rm{loc}}(C,A):=
        \{\tr\,(C^\dagger KAK^\dagger) \;|\; K \in SU_{\rm loc}(2^n)\} \subseteq W(C,A).
\end{equation}
It can be viewed as a projection of the local unitary orbit of the initial state $A$
onto the target state $C$.
\end{definition}

As also seen in the concomitant study, in contrast to the usual $C$"~numerical range,
its local counterpart is no longer star-shaped, nor simply connected.
With these stipulations, we will discuss recent applications of
the local $C$"~numerical range in quantum control.

\subsection{Application in Quantum Information}

Again, in terms of Euclidean geometry, maximising the real part in $W_{\rm loc}(C,A)$ 
minimises the distance from $C$ to the local unitary orbit
$\mathcal O_{\rm loc} (A)$.
%$$
%\maxover {K\in SU(2)^{\otimes n}} \Re\tr\{C^\dagger \,KAK^{-1}\} \Leftrightarrow
%\minover {K\in SU(2)^{\otimes n}} \fnorm{KAK^{-1} - C}
%$$

In Quantum Information Theory, the minimal distance has an interesting interpretation
in the following setting: let $A$ be an arbitrary rank-$1$ state of the form $A = \ketbra{\psi}{\psi}$ 
and let $C=\diag(1,0,\dots,0)\in\Mat_{2^n}(\C{})$. Thus in this case 
$W_{\rm loc}(C,A)$ reduces to what we define as
the {\em local field of values} ${W_{\rm loc}(A)}$.

\begin{definition}[(Pure-State Entanglement)]\\
An $n$"~qubit pure state $A=\ketbra \psi \psi$ with $\ket\psi\in\C{2^n}$ is termed a {\em product state}, if it can be written
as a tensor product
\begin{equation}
A = \ketbra{\psi_1}{\psi_1}\otimes\ketbra{\psi_2}{\psi_2}\otimes\cdots\otimes\ketbra{\psi_n}{\psi_n}\quad\text{with}\quad \ket{\psi_j}\in\C 2\quad,
\end{equation}
whereas it is said to be {\em entangled} if it cannot.
\end{definition}
\begin{remark*} In the present context, there are important observations with regard to the full unitary
        orbit $\mathcal O_u(A)$ and the {\em local} unitary orbit
$\mathcal O_{\rm loc}(A)$ of pure states $A$
        of different types:\\
\begin{enumerate}
\item[1.] all pure states form an equivalence class coinciding with
$\mathcal O_u(A)$ if $A=\ketbra\psi\psi$ is
        an arbitrary pure state;
\item[2.] generic elements on the full unitary orbit $\mathcal O_u(A)$ of a pure product state $A$
        are pure, yet no longer of product form;
\item[3.] all pure product states form an equivalence class coinciding with the {\em local}
        unitary orbit $\mathcal O_{\rm loc}(A)$ of an arbitrary pure product state $A$.\\
%\item[4.] consequently, measures of entanglement remain invariant under {\em local} unitary transformation.
\end{enumerate}

\noindent
Consequently, measures of entanglement remain invariant under {\em local} unitary transformation.
\end{remark*}

\begin{corollary}[(Euclidean Measure of Pure-State Entanglement)]\\
For $A = \ketbra{\psi}{\psi}$, $C=\diag(1,0,\dots,0)\in\Mat_{2^n}(\C{})$
%and $\mathbf K := SU_{\rm loc}(2^n)$,
the minimial Euclidean distance 
\begin{equation}
\Delta:=\minover{K \in SU_{\rm loc}(2^n)}\; ||KAK^\dagger - C||_2
\end{equation}
is a {measure of pure-state entanglement} because
it quantifies how far $A$ is from the {equivalence class} of
pure product states.
It relates to the maximum real part of the local numerical range $W_{\rm loc}(A)$ via
\begin{equation}
\begin{split}
  {|| C - KAK^\dagger ||}_2^2 &= {||A||}_2^2 + {||C||}_2^2 - 2 \Re \tr \{C^\dagger\; K A K^\dagger\}\\
                              &=  2 - 2 \Re \tr \{C^\dagger\; K A K^\dagger\}\quad,  
\end{split}
\end{equation}
where the last equality holds if also $A$ is normalised to $||A||_2=1$. 
\end{corollary}

\begin{figure}
\begin{center}
\includegraphics[height=42mm]{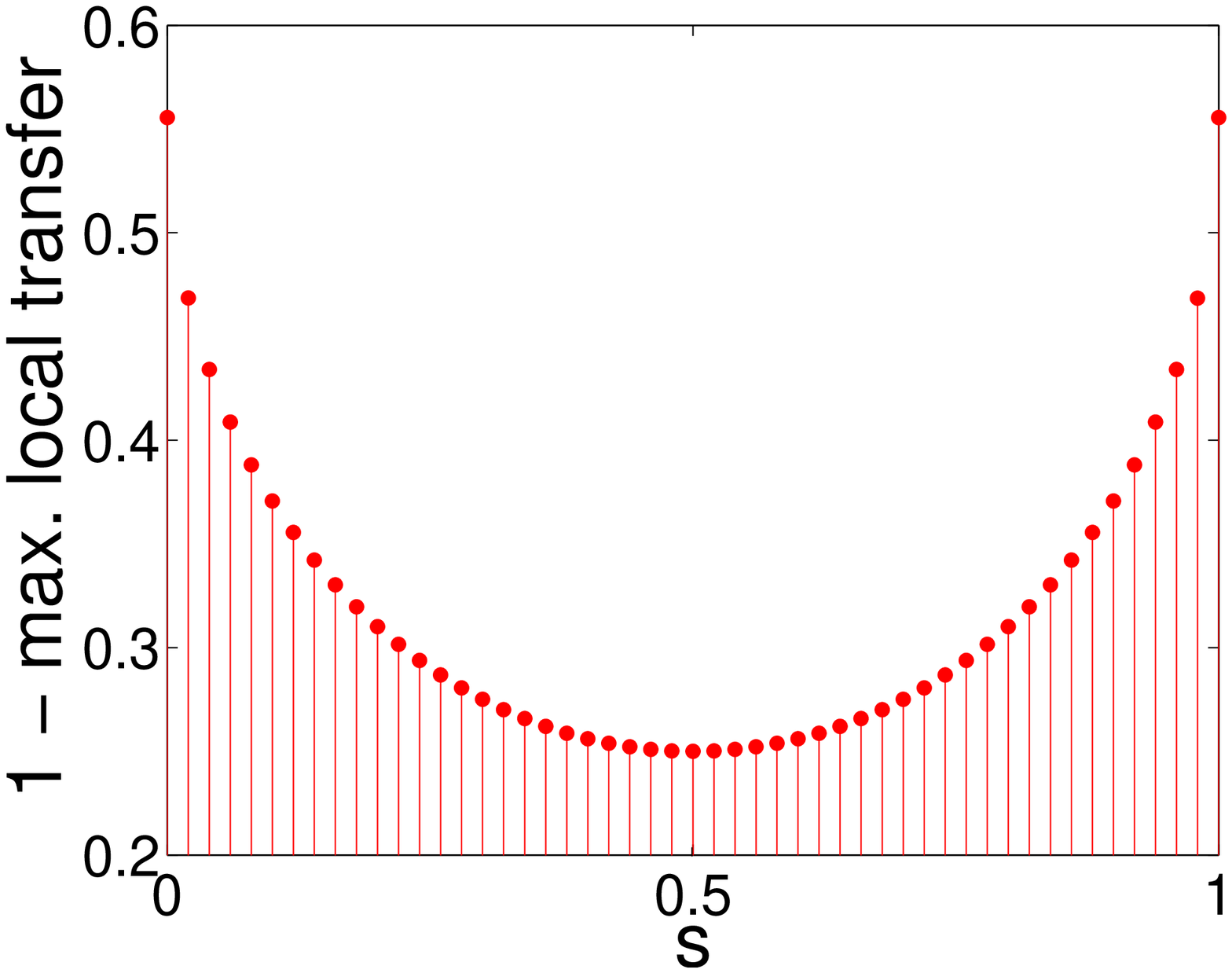}
\hspace{3mm}
\includegraphics[height=42mm]{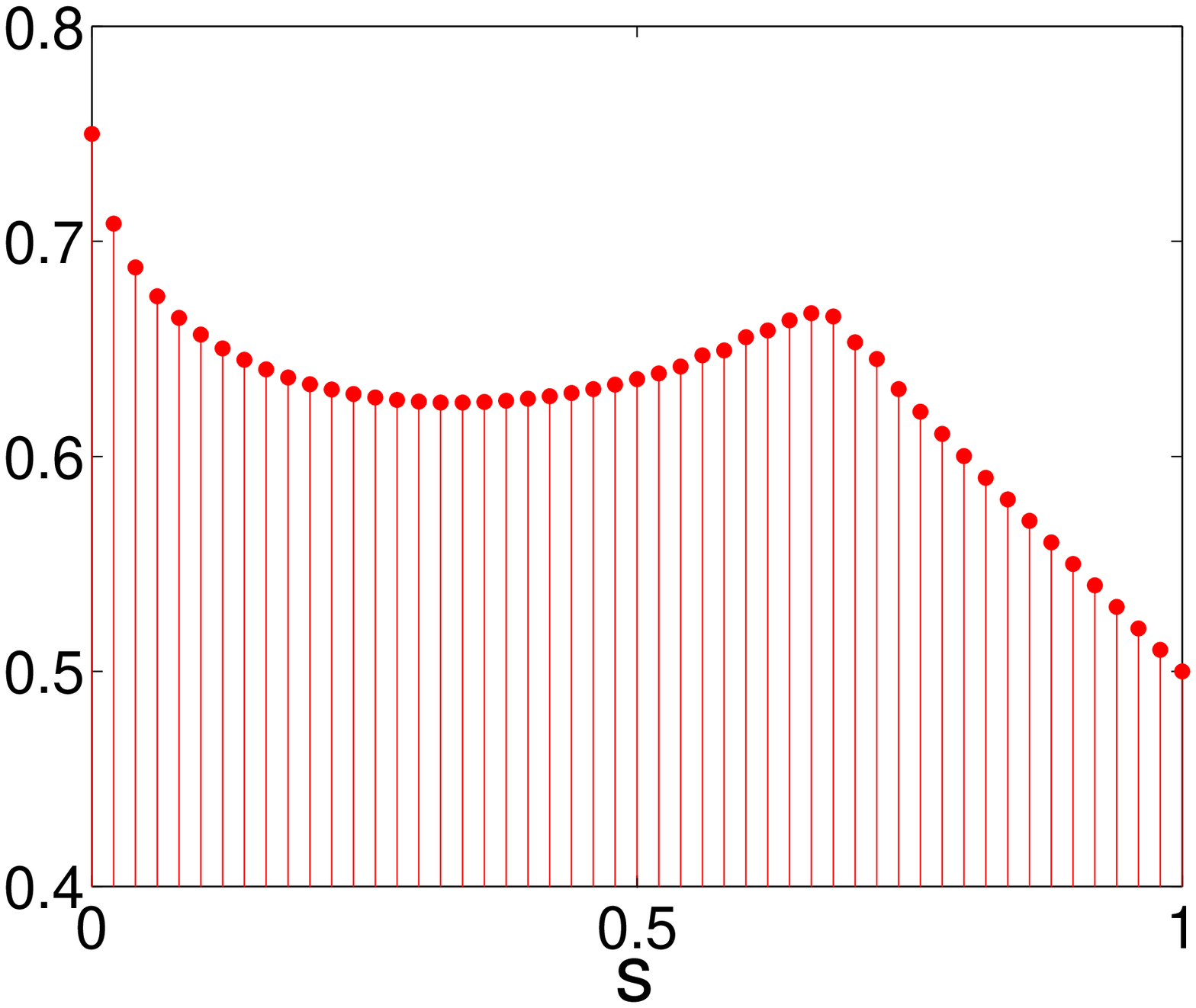}
\end{center}
%\label{fig:guehne}
\caption{\label{fig:guehne}
Euclidean distances of the pure states of Example~\ref{ex:3qubit} (a) and Example~\ref{ex:4qubit} (b)
to the nearest pure product state as a function of $s$;
here \/`max. local transfer\/' denotes $\Re \tr \{C^\dagger\; K A K^\dagger\}$.
}
\end{figure}

The (squared) Euclidean distance from the nearest pure product state is illustrated in Fig.~\ref{fig:guehne}
for the following two examples taken from quantum information theory \cite{Guehne04,Goldbart03}:

%%%
%%%
\begin{example}\label{ex:3qubit}
First, consider the pure 3-qubit state $A(s):=\ketbra{\psi_3(s)}{\psi_3(s)}$ parameterised by $0\leq s \leq 1$
\begin{equation}\label{psi1}
|\psi_3(s)\rangle := \sqrt{s} |W\rangle + \sqrt{1-s} |\tilde{W}\rangle\quad,% \text{with}\quad 0\leq s \leq 1\quad, 
\end{equation}
where\;
$\ket W := \tfrac{1}{\sqrt{3}} {(0,1,1,0, 1,0,0,0)^t}\quad{\rm and}\quad
\ket{\tilde W} := \tfrac{1}{\sqrt{3}}{(0,0,0,1, 0,1,1,0)^t}$\quad .
\end{example}
%%%%%%%%%%%%%%%%%%%
%%%%%%%%%%%%%%%%%%%

\begin{example}\label{ex:4qubit}
Likewise, observe the pure 4-qubit-state $A(s):=\ketbra{\psi_4(s)}{\psi_4(s)}$
\begin{equation}\label{psi2}
\ket{\psi_4(s)}:=\sqrt{s}\ket{GHZ'} - \sqrt{1-s}\ket{\psi^+}\otimes\ket{\psi^+}\quad,
\end{equation}
where\quad
$\ket{\rm GHZ'} := \tfrac{1}{\sqrt{2}} {(0,0,0,1, 0,0,0,0, 0,0,0,0, 1,0,0,0)^t}$\quad{\rm and}\\
$\ket{\psi^+}\otimes\ket{\psi^+} := \tfrac{1}{2} {(0,0,0,0, 0,1,1,0, 0,1,1,0, 0,0,0,0)^t}$\quad .
\end{example}
\noindent
Note that the plots in Fig.~\ref{fig:guehne} quantitatively reproduce the global optimisation
results found by numerical quadratic programming methods \cite{Guehne04} yet cutting the 
{\sc cpu} time by a factor of approx.~$35$ in Example~\ref{ex:3qubit} and by a factor of 
some $150$ in Example~\ref{ex:4qubit}. Moreover, in both cases,
the findings coincide with the exact solutions known algebraically \cite{Goldbart03}.

%%%
\noindent
{\em Caveat}: in non-pure states, the problem of  entanglement is much more involved,
since the state space to non-pure states forms no simplex: a generic density operator has
infinitely many decompositions into pure states. For recent overviews, 
see, e.g., ref.~\cite{BenZyc06,BruLeu07}.
Hence in those instances the above approach no longer applies.
%%%
\subsubsection{Significance of Entanglement}
Finally, it is the purpose of this tutorial paragraph to show the numerical-range focussed reader
why in {\em entangled} quantum systems,
the total system comprises more information than accessible from putting together the information of
all its {\em local} subsystems.

%To this end, we want to express a bipartite system $(a,b)$ in terms of its subsystems $a$ and $b$
%by making use of the respective orthonormal Hilbert space vectors $\ket{a_\nu}\in\HS_a$ and $\ket{b_\mu}\in\HS_b$.
%Then a generic density operator in the total Hilbert space $\mathcal H_a \otimes \mathcal H_b$
%can be expanded in the Schmidt bases as
To this end, we want to express a bipartite system $(a,b)$ in terms of its subsystems $a$ and $b$
by making use of the respective orthonormal Hilbert space vectors $\ket{a_\nu}\in\HS_a$ and 
$\ket{b_\mu}\in\HS_b$. For simple demonstration, we assume both to be of the same dimension
and consider a type of density operator in the total Hilbert space 
$\mathcal H_a \otimes \mathcal H_b$ that can be expanded in the Schmidt bases as
\begin{equation}
        \rho_{ab}=\sum_{\mu\nu}\lambda_{\mu\nu}(\ketbra{a_\mu}{a_\nu}\otimes\ketbra{b_\mu}{b_\nu})\quad.
\end{equation}
The information locally accessible in subsystem $a$ is encoded in the {\em reduced density operator}
of subsystem $a$ that is projected out by taking the so-called {\em partial trace} 
$\tr(\ketbra{b_\mu}{b_\nu})$
over the degrees of freedom of subsystem $b$ yielding
        \begin{equation}
        \begin{split}
        \rho_a := \tr_b\{\rho_{ab}\}
               &= \tr_b\{\Sigma_{\mu\nu}\lambda_{\mu\nu}(\ketbra{a_\mu}{a_\nu}\otimes\ketbra{b_\mu}{b_\nu})\} \\
               &= \Sigma_{\mu\nu}\lambda_{\mu\nu}(\ketbra{a_\mu}{a_\nu}\cdot\tr\{\ketbra{b_\mu}{b_\nu}\})\\
               &= \Sigma_{\mu\nu}\lambda_{\mu\nu}(\ketbra{a_\mu}{a_\nu}\cdot{\braket{b_\nu}{b_\mu}}) \\
               &= \Sigma_{\mu}\lambda_{\mu\mu}\ketbra{a_\mu}{a_\mu} \quad,
        \end{split}
        \end{equation}
where the last equality holds, because in the orthonormal Schmidt basis $\braket{b_\nu}{b_\mu}=\delta_{\mu\nu}$.
The following standard examples \cite{NC00}
will illustrate reduced states in the scenarios of product states on one hand,
and entangled states on the other.
\begin{example}
Product states take the form $\rho_{ab} = \rho_a\otimes\rho_b$, so one trivially finds
\begin{equation}
        \rho_a    = \tr_b\{\rho_a\otimes\rho_b\} = \rho_a\cdot\tr\rho_b = \rho_a
\end{equation}
since $\tr\rho_b =1$ by definition of a normalised density operator.

\end{example}

\begin{example}
However, in the Bell state $\ket{\Phi}:=\tfrac{1}{\sqrt{2}}(\ket{00}+\ket{11})$, 
where it is customary to use the short-hand
$\ket{0}:=\left(\begin{smallmatrix}1\\0\end{smallmatrix}\right)\in\C 2$
        and $\ket{1}:=\left(\begin{smallmatrix}0\\1\end{smallmatrix}\right)$ as well as 
        $\ket{00}:=\ket 0 \otimes \ket 0\in\C 4$ and likewise $\ket{11}:=\ket 1 \otimes \ket 1$,
one obtains
\begin{equation}
        \begin{split}
        \rho_{ab} &= \ketbra{\Phi}{\Phi} = \tfrac{1}{2}(\ket{00}+\ket{11})(\bra{00}+\bra{11})\\[2mm]
        \rho_a    &= \tr_b\ketbra{\Phi}{\Phi}\\
                  &= \tfrac{1}{2}\tr_b\big(\ketbra{00}{00}+\ketbra{00}{11}+\ketbra{11}{00}+\ketbra{11}{11}\big)\\
                  &= \tfrac{1}{2}\big(\ketbra{0}{0}_a{\braket{0}{0}_b}+\ketbra{0}{1}_a{\braket{1}{0}_b}%
                        +\ketbra{1}{0}_a{\braket{0}{1}_b}+\ketbra{1}{1}_a{\braket{1}{1}_b}\big)\\
                  &= \tfrac{1}{2}\big(\ketbra{0}{0}_a+\ketbra{1}{1}_a\big)\\
                  &= \tfrac{1}{2}\unity_a\\[2mm]
        \rho_b    &= \tr_a\ketbra{\Phi}{\Phi} = \tfrac{1}{2}\unity_b
        \end{split}
        \end{equation}
The second example shows the generic situation: although $\rho_{ab}$ is a pure state, the reduced
states of the respective subsystems $\rho_{a(b)}\in\HS_{a(b)}$ are no longer pure.
\end{example}

\begin{remark*}
{Compare the information content}\/:
\begin{enumerate}
\item[(1)] $\rho_{ab}^{\phantom{2}}=\rho_{ab}^2$ contains {\em all information} about the total system;
\item[(2)] $\rho_{a(b)}$ contains all information about the respective subsystem $a(b)$;
\item[(3)] the reconstruction 
        $\rho_{ab}':=\rho_a\otimes\rho_b \neq \rho_{ab}'^2$ puts together {\em all information accessible
        from both local subsystems}; this is generically less than in $\rho_{ab}$.\\
\end{enumerate}
In the second example, the reconstruction
$\rho_{ab}'= \rho_a \otimes \rho_b = \tfrac{1}{4}\unity$ 
is {\em diagonal}, whereas the original $\rho_{ab} = \sum_{\mu\nu} \lambda_{\mu \nu}
                (\ketbra{a_\mu}{a_\nu}\otimes\ketbra{b_\mu}{b_\nu})$ contained {\em off-diagonal} terms.
Thus it is the coherent phase relation between the local constituents that is inevitably lost upon projection to 
the respective reduced systems. It cannot be reconstructed {\em a posteriori} using but {\em local} pieces of information.
\end{remark*}
This shows how in {\em entangled} quantum systems,
the total system comprises more information than is accessible from putting together the information of
all its reduced {\em local} subsystems. Measures of entanglement account for this loss of information
and thus play an important role in quantum information theory.

%%%
%%%

\subsection{Application in Quantum Simulation}
In quantum control, it is of interest to decide, whether a given 
multi-particle quantum interaction (expressed by some interaction Hamiltonian $H$)
can be sign-reversed solely by local unitary operations. If this is the case,
one may undo or refocus the time evolution of such an interaction purely by {\em local operations}
generalising the sense of the celebrated Hahn spin echo \cite{Hahn50,EBW87} in the following way: 
(0) start with any initial sate, (1) let the interaction evolve for some time $t$ 
to give the propagator $e^{-itH}$, (2) apply appropriate local operations,
(3) let the interaction evolve again for the same duration $t$, (4) apply the inverse to the previous
local operations to (5) recover the initial state again as an echo, 
because steps (2)--(3)--(4) bring about the inverse propagator $e^{+itH}$.
Note the same local operations apply to all the initial states; they only depend on the
interaction Hamiltonian $H$.

Mathematically, we ask whether there is a $K \in SU_{\rm loc}(2^n)$ such that
\begin{equation}
                K\; e^{-it H}\; K^\dagger = e^{+it H}\quad
\mbox{for all}\; t\in \R+\quad.
\end{equation}
Due to the equality $K\; e^{-it H}\; K^\dagger = e^{-it KHK^\dagger}$
%series expansion of the exponential
the question readily boils down to deciding whether the sign-reversed
Hamiltonian $-H$ is on  the {\em local unitary orbit} of the original
Hamiltonian $H$, i.e., does there exist $K \in SU_{\rm loc}(2^n)$
such that
\begin{equation}
                K\; H\; K^\dagger = - H\quad.
\end{equation}
Recently, we have solved this problem based on its normal form \cite{PRA_inv}: this
is sign reversibility by local $z$"~rotations,
since every element $K \in SU_{\rm loc}(2^n)$ is locally unitarily similar to local
$z$"~rotations.
Here, we focus on the relation to local $C$"~numerical ranges
%$W_{\rm loc}(C,A)$
by illustrating that normalised Hamiltonians $H$
% $H$ normalised to $||H||_2=1$,
are sign-reversible if and only if $W_{\rm loc}(H,H) = [-1;+1]$, cf. \cite{PRA_inv}.
%%%%%%%%%%%%%%%%%%%%
%%%%%%%%%%%%%%%%%%%%
%\noindent
To this end, define the spin-$\tfrac{1}{2}$ operators
\begin{equation}
J_0:=\unity_2\;;\quad
J_z:=\tfrac{1}{2}\diag(1,-1)\;;\quad
J_+:= \left(\begin{smallmatrix} 0 &1\\0 &0\end{smallmatrix}\right)\;;\quad
J_-:= J_+^\dagger = \left(\begin{smallmatrix} 0 &0\\1 &0\end{smallmatrix}\right)\;.
\end{equation}
%%%%%%
In a single qubit, the $J_\nu \in \{J_0, J_z, J_+, J_-\}$ are the
eigenoperators to the conjugation map $\Ad {e^{-i\phi J_z}}$, i.e. 
\begin{equation}
\Ad {e^{-i\phi J_z}}(J_\nu):= e^{-i\phi J_z}\;J_\nu\;e^{+i\phi J_z} = e^{-i p_\nu \phi} \;J_\nu 
\end{equation}
associated with the respective eigenvalues
$e^{-i p_\nu \phi}$,
%\in \{1,1,e^{-p_+\phi},e^{-p_-\phi}\}$
$p_0=p_z=0$ and $p_\pm=\pm1$.
In order to generalise the arguments to the case of $z$"~rotations
on $n$ qubits with individually differing rotation angles on each
spin qubit $\phi_1, \phi_2, \dots, \phi_\ell, \dots, \phi_n$,
we write
\begin{equation}
K_z(\phi_1,\; \dots\;, \phi_n) =
e^{-i\phi_1 J_z} \otimes e^{-i\phi_2 J_z}
\otimes \dots \otimes e^{-i\phi_n J_z} \in {SU}_{\rm loc}(2^n).
\end{equation}
Now consider a Hamiltonian in normal form $\bar H$ taking the special form
$\bar H :=  \bar H_+ + \bar H_-$ with $\bar H_- := \bar H_+^\dagger$, where $\bar H_+$ is
a tensor product of $\Ad {e^{-i\phi J_z}}$ eigenoperators on each spin qubit
$\ell=1,\dots, n$
according to
\begin{equation}
\label{eqn:ham_plus}
\bar H_+ := \JZ 1 \otimes \JZ 2 \otimes \cdots \otimes \JZ \ell \otimes \cdots \otimes \JZ n
\end{equation}
with independent $\nu_\ell \in \{0, z, +, -\}$
on each spin qubit. With $\bar H_\pm$ satisfying
%being composed of eigenoperators to individual local $z$-rotations 
\begin{equation}
\Ad {K_z(\phi_1,\; \dots\;, \phi_n)} \big(\bar H_\pm \big) = %
   e^{\mp i (p_{\nu_1} \phi_1 +\; \cdots \;+ p_{\nu_n} \phi_n)} \; \bar H_\pm\;,
\end{equation}
the Hamiltonian $\bar H$ is sign-reversed by local $z$"~rotations provided there exists a set of rotation
angles $\{\phi_\ell\}$ satisfying
$
\sum_{\ell = 1}^n p_{\nu_\ell} \phi_\ell = \pm \pi\; (\mod 2 \pi)\; .
$
This is the case if there is at least one spin qubit $\ell$ giving rise to an
interaction of quantum order $p_{\nu_\ell} = \pm 1$. 

\medskip
Moreover, a (real) linear combination
\begin{equation}\label{eqn:ham_sigma}
\bar H_\Sigma := \sum_{\lambda=1}^m c_\lambda \bar H_\lambda
\end{equation}
of Hamiltonians $\bar H_\lambda := \bar H_{\lambda +} + \bar H_{\lambda -}$ with
$\bar H_{\lambda +}$ as in Eqn.~\ref{eqn:ham_plus}
%$H_\Sigma := \sum_{\lambda=1}^m c_\lambda H_\lambda$
is jointly reversible by individual local $z$"~rotations
$K_z(\phi_1,\; \dots\;, \phi_n)$ ,
if there is at least one consistent
set of rotation angles $\{\phi_\ell\}$ simultaneously fulfilling
a standard linear system of $m$ equations in $n$ variables
\begin{equation}\label{eqn:lin_syst}
\begin{pmatrix}
        p_{1\nu_1} & p_{1\nu_2} & \cdots & p_{1\nu_n} \\
        p_{2\nu_1} & p_{2\nu_2} & \cdots & p_{2\nu_n} \\
        p_{3\nu_1} & p_{3\nu_2} & \cdots & p_{3\nu_n} \\
        \vdots & \vdots & \ddots & \vdots \\
        p_{m\nu_1} & p_{m\nu_2} & \cdots & p_{m\nu_n} \\
\end{pmatrix}
\begin{pmatrix}\phi_1\\ \phi_2\\ \vdots\\ \phi_n\end{pmatrix} =
\begin{pmatrix} \pi\; (\mod 2 \pi)\\  \pi\; (\mod 2 \pi)\\  \pi\; (\mod 2 \pi)\\ \vdots\\ %
                 \pi\; (\mod 2 \pi)\end{pmatrix}.
\end{equation}
%%%%%%%%%%%%%%%%%%%%
%%%%%%%%%%%%%%%%%%%%

With these stipulations, we have recently proven the following 
interrelations in view of local $C$-numerical ranges:
%%%%%%%%
%\newpage
%%%%%%%%

\begin{corollary}[(Local Time Reversal \cite{PRA_inv})]\label{cor:loc_inv}\\[2mm]
For $H=H^\dagger$ with $\fnorm H = 1$ the following are equivalent:
\begin{enumerate}
\item[(1)] the Hamiltonian $H$ is sign-reversible under local unitary operations;
\vspace{2mm}
\item[(2)] its {local $C$-numerical range} comprises $-1$, 
                i.e., $ -1 \in W_{\rm loc}(H,H)$;
\vspace{2mm}
\item[(3)] its {local $C$-numerical range} is the interval $[-1\,;\,+1]\;=\;W_{\rm loc}(H,H)$;
\vspace{2mm}
\item[(4)] there exists a $K\in SU_{\rm loc}(2^n)$
such that
%$\fnormsq{KHK^{-1} + H} = 0$ \;or equivalently :\;
$\Adr_K(H) = -H$;
\vspace{2mm}
\item[(5)] $H$ is locally unitarily similar to a $\Bar{H\;}$ with
$\Adr_{K_z(\phi_1,\dots, \phi_n)}(\Bar{H\;}) = -\Bar{H\;}$;
\vspace{2mm}
\item[(6)] $H$ is locally unitarily similar to a linear combination of the form
	Eqn.~\ref{eqn:ham_sigma} satisfying the system of linear equations given 
	in Eqn.~\ref{eqn:lin_syst};
\vspace{2mm}
\item[(7)]
let $\mathfrak g = \mathfrak g_0 \oplus\;\bigoplus\limits_{i\neq j} \C{} E_{ij}$
be the root-space decomposition of $\mathfrak{sl}(N,\C{})$, where $E_{ij}$
denotes the square matrix differing from the zero matrix by just
one element, the unity in the $j^{\rm th}$ column of the $i^{\rm th}$ row;
then $H$ is locally unitarily similar to a linear combination of 
root-space elements to non-zero roots (so $i\neq j$) with
$\Bar H_\Sigma\,  := \sum_{\lambda=1}^m c_\lambda E_{ij}^{(\lambda)}$ 
satisfying the system of linear equations
$\sum_{\ell=1}^n p_{\lambda,\ell}\cdot\phi_\ell = \pi(\mod 2 \pi)$
for $\lambda=1,2,\dots, m$ as in Eqn.~\ref{eqn:lin_syst}.
\end{enumerate}
\end{corollary}

\noindent
In ref.~\cite{PRA_inv}, we provided more tools to assess local
reversibility by means of eigenspaces, graph representations of the interaction topology,
spherical tensor methods, and root-space decomposition.
Based on assertion (4) we also implemented a gradient-flow algorithm 
on the group of local unitaries $SU_{\rm loc}(2^n)$ in
order to tackle the problem numerically.

In the accompanying paper, we show the following:
%%%%%%%%
%%%%%%%%
%%%%%%%%
\begin{theorem}  [(Local $C$"~Numerical Ranges of Circular Disc Shape \cite{LAMA_WKCA})]\label{thm:circ_wkca}\\
Let $\mathbf K$ be a compact connected subgroup of $U(N)$ with Lie algebra $\mathfrak k$,
and let $\mathfrak t$ be a torus algebra of $\mathfrak k$. Then the relative 
$C$"~numerical range $W_{\mathbf K}(C,A_+)$ of a matrix
$A_+\in\Mat_N(\C{})\setminus 0$ is a circular
disc centered at the origin of the complex plane for all $C\in\Mat_N(\C{})$ if and only if
%\begin{enumerate}
%\item[(1)] there exists an $\Omega \in \mathfrak k$ so that $A$ is an eigenoperator to $\adr_\Omega$
%       with a non-zero eigenvalue
%       $$\adr_\Omega(A) = \lambda A \quad\text{\rm and}\quad \lambda\neq 0$$
%\end{enumerate}
%or equivalently
%\begin{enumerate}
%\item[(1)] 
        there exists a $K\in\mathbf K$ and a $\Delta\in\mathfrak t$ such that
        $KA_+K^\dagger$ is an eigenoperator to $\adr_\Delta$ with a non-zero eigenvalue
        \begin{equation}\label{eq:ksymm}
        \adr_\Delta(KA_+K^\dagger) \equiv \comm{\Delta}{KA_+K^\dagger} 
                = i p \;KA_+K^\dagger \quad\text{\rm and}\quad p \neq 0\quad.
        \end{equation}
Clearly, if $K A_+ K^\dagger$ is an eigenoperator of $\adr_\Delta$ to the
eigenvalue $+ip$
and $A_-:=A_+^\dagger$, then $K A_- K^\dagger$ shows the eigenvalue $-ip$. 
$A_+$ and $A_-$ share the same
relative $C$"~numerical range of circular symmetry, $W_K(C,A_+)=W_K(C,A_-)$.
\end{theorem}

Moreover, locally reversible Hamiltonians $H$ and nilpotent matrices $\{A_+, A_-\}$ with rotationally
symmetric local $C$"~numerical ranges are related as follows:

\begin{corollary}\\
Let $\mathbf K = SU_{\rm loc}(2^n)$. Let $A_+$ and $A_-:=A_+^\dagger$ both share the same
local $C$"~numerical ranges $W_{\rm loc}(C,A_\pm)$ of circular disc shape for all $C$.\\
Then 
\begin{enumerate}
\item[(1)] any linear combination $A_\lambda := A_+ + \lambda A_-$ with $\lambda \in \C{}$
        and in particular the Hermitian $H:= A_+ + A_-$ are sign reversible by some local $K\in SU_{\rm loc}(2^n)$,
        so $W_{\rm loc}(H,H) = [-||A||_2^2; + ||A||_2^2]$, whereas
\item[(2)] the converse does not necessarily hold: there are locally reversible Hermitian $H$
        to which no decomposition into a single pair $\{H_+,H_-\}$ sharing the same
        rotationally symmetric local $C$"~numerical range $W_{\rm loc}(C,H_\pm)$ exist, but
\item[(3)] every Hermitian $H\in\Mat_{2^n}(\C{})$ that is locally sign reversible can be decomposed into at most
        ${2^n}\choose{2}$ pairs
$(H^{(1)}_+,H^{(1)}_-), (H^{(2)}_+,H^{(2)}_-), \dots$ 
        with each pair sharing the same rotationally symmetric local $C$"~numerical range 
        $W_{\rm loc}(C,H^{(\ell)}_\pm)$.
\end{enumerate}
\end{corollary}

\begin{proof}:
\begin{enumerate}
\item[(1)] Eqn.~\ref{eq:ksymm} is equivalent to\vspace{-3mm}
        \begin{equation}
	\begin{split}
        \Adr_{e^{-\phi\Delta}}(KA_+K^\dagger) &\equiv e^{-\phi\Delta} (KA_+K^\dagger) e^{+\phi\Delta} \\
                &= e^{-\phi\adr_\Delta} (KA_+K^\dagger)= e^{-i\phi p} KA_+K^\dagger\quad,
	\vspace{-9mm}
	\end{split}
        \end{equation}
        thus $p\neq 0$ ensures a $\phi$ with $e^{\pm i\phi p}=-1$ to sign-reverse both
        $KA_\pm K^\dagger$.
\item[(2)] By Corollary~\ref{cor:loc_inv}, local sign-reversibility allows for linear 
        combinations of eigenoperators to {\em different}
        $\adr_\Delta$-operators with different eigenvalues thus generically
        violating the conditions for rotational symmetry of Theorem~\ref{thm:circ_wkca}.
\item[(3)] By Corollary \ref{cor:loc_inv} any Hermitian
$\bar H\in\Mat_{2^n}(\C{})$ locally reversible by $z$"~rotations 
        can trivially be decomposed into 
        at most $2^n\choose 2$ Hermitian components 
        $(\lambda_{ij}E_{ij}+\lambda_{ij}^*E_{ji})$ with $1\leq i<j\leq N=2^n$
        and $\lambda_{ij}\in\C{}$, where the $(\bar H_+^{(\ell)},\bar H_-^{(\ell)}):=
        (\lambda_{ij}E_{ij}, \lambda_{ij}^*E_{ji})$
share the same rotationally symmetric local $C$"~numerical range.
\end{enumerate}
\vspace{-6mm}
\end{proof}
%%%%%%%%%%%%%%%%%%%%
%%%%%%%%%%%%%%%%%%%%

\section{Constrained Optimisation and Relative $C$"~Numerical Ranges}\label{chp:const_WCA}
In quantum control, one may face the problem to maximise the unitary
transfer from matrices from $A$ to $C$ subject to suppressing the transfer
from $A$ to $D$,
or subject to leaving another state $E$ invariant.
\noindent
For tackling those types of problems, in ref.~\cite{TOSH-Diss}
we introduced a \/`constrained $C$-numerical range of $A$\/'.

\begin{definition}\\% [(Constrained $C$-Numerical Range of $A$)]\\
The {\em constrained $C$-numerical range of $A$} is defined by
\begin{equation}
W(C,A)\big|_{\rm constraint}\,:=\, \big\{\tr(UA\Ud \Cd)\,\big|\; \text{constraint} \big\} \subseteq \wca\;.
\end{equation}
\end{definition}
\noindent
In ref.~\cite{TOSH-Diss} we also asked which form it takes and---in view of numerical optimisation---whether it is a 
connected set with a well-defined boundary.
% $\partial{W(C,A)|_{\rm constraint}}$.
Connectedness is central
to any numerical optimisation approach, because otherwise one would have to rely on initial conditions in
the connected component of the (global) optimum.

\medskip
Exploiting the findings on the relative $C$"~numerical range of the accompanying paper \cite{LAMA_WKCA}, 
the structure of constrained $C$"~numerical ranges can readily be characterised in some
simple cases.

\begin{corollary}\\
The constrained $C$"~numerical range of $A$ is a connected set in the
complex plane, if the constraint can be fulfilled by restricting the full
unitary group $U(N)$ to a compact and connected subgroup $\mathbf K \subseteq U(N)$.

\noindent
In this case, the constrained $C$"~numerical range $\wca|_{\rm constraint}$
is identical to the relative $C$"~numerical range $W_{\mathbf K}(C,A)$
and hence the constrained optimisation problem 
is solved within it, e.g.,
by the corresponding relative $C$"~numerical radius $r_{\mathbf K}(C,A)$\,.
\end{corollary}

\begin{proof}:
Direct consequence of the properties of the relative $C$"~numerical
range \wkca introduced in the accompanying paper \cite{LAMA_WKCA}: if $\mathbf K$ is compact
and connected, then \wkca is so as well, since
it is a continuous image of a compact and connected set.
\end{proof}

\begin{remark*}
Note that although being connected, \wkca is in general neither star-shaped nor simply
connected \cite{LAMA_WKCA}. So if $\mathbf K$ is compact and connected
this obviously extends to $\wca|_{\rm constraint}$.
\end{remark*}
%%%%%
%%%%%
%%%%%%%%%%%%%%%%%%%%%%%%%%%%%%%
\subsubsection{Constraint by Invariance}
The problem of maximising the transfer from $A$ to $C$ while leaving $E$ invariant
%%%%%
\begin{equation}\label{eqn:inv_copt}
\maxover U |\tr\{UA\Ud \Cd\}|\quad\text{\rm subject to}\quad UE\Ud= E 
\end{equation}
%%%%%
is straightforward in as much as the stabiliser group of $E$ 
%%%%%
\begin{equation}\label{eq:stab_E}
\mathbf K_E := \{ K\in U(N) \,|\, KEK^\dagger = E \}
\end{equation}
%%%%%
is easy to come by: it is generated by the Lie-algebra elements
%%%%%
\begin{equation}\label{eq:stab_E2}
\mathfrak k_E := \{ k\in \mathfrak{u}(N) \,|\, \adr_{k}(E)\equiv \comm{k}{E} = 0 \}\quad.
\end{equation}
In particular, %other words
if $E$ is of the form $E=\mu\unity + \Omega$ with $\mu\in\C{}$ and $\Omega\in\mathfrak{u}(N)$, 
then $\mathfrak k_E$ is identical to the centraliser of $\Omega$ in $\mathfrak{u}(N)$.
%%%%%
\begin{lemma}\\
The set $\mathfrak k_E$ is closed under the Lie bracket, hence it is a
subalgebra to $\mathfrak{u}(N)$ thus generating a subgroup $\mathbf K_E \subseteq U(N)$, to wit the stabiliser group.
\end{lemma}
%%%%%
\begin{proof}:
Direct consequence of the Jacobi identity for the double commutator: 
$[[A,B],E] + [[B,E],A] + [[E,A],B] = 0$ for all $A,B,E \in \Mat_N(\C{})$. Hence $[k_1,E]=[k_2,E]=0$ implies $[[k_1,k_2],E]=0$
and $\mathfrak k_E$ is a Lie subalgebra to $\mathfrak{u}(N)$ thus generating a compact connected stabiliser
group $\mathbf K_E \subseteq U(N)$.
\end{proof}
%%%%%

\begin{remark*}
The stabiliser group of any $E \in \Mat_N(\C{})$ in $U(N)$ is connected.
This is in general not the case in $SU(N)$ as easily seen for
$E := \left( \begin{smallmatrix} 0 & 1\\0 & 0 \end{smallmatrix}\right)$.
However, one can restrict the above optimisation to the connected component
of the identity matrix in SU(N) due to the invariance properties of the
function $U \mapsto \tr\{UA\Ud \Cd\}$.
\end{remark*}

%Let $\langle k_j | j=1,2,\dots,\dim\big(\mathfrak{su}(N)\big)\rangle_{\rm Lie} = \mathfrak{su}(N)$.
%So $\{k_j|j=1,2,\dots,N^2-1\}$ denotes
%the set of generators forming $\mathfrak{su}(N)$ by way of commutation.
%In the general case,
A set of generators of $\mathfrak k_E$ may constructively be found via
the kernel of the commutator map by solving a homogeneous linear system
\begin{equation}\label{eqn:centraliser}
\mathfrak k_E = \ker\adr_E \cap \,\mathfrak{su}(N) = 
        \{ k \in \mathfrak{su}(N) | (\unity\otimes E - E^t\otimes\unity) \vec(k) = 0\}\quad.
\end{equation}

\begin{corollary}\\
If solvable in an non-trivial way, the optimisation problem
of Eqn.~\ref{eqn:inv_copt} entails a constrained $C$"~numerical range 
that is connected. Moreover, it takes the form of a relative $C$"~numerical range
\begin{equation}
\wca\big|_{\Adr_U(E)=E}:=\{\tr(UA\Ud\Cd)\,|\,UE\Ud = E\} = W_{\mathbf K_E}(C,A)\quad
\end{equation}
and the optimisation problem is solved by the relative $C$"~numerical radius $r_{\mathbf K_E}(C,A)$.
In Hermitian $E$, $\mathbf K_E$ includes a maximal torus group $\mathbf T \subset SU(N)$.
\end{corollary}
\begin{proof}: To any $E\in\Mat_N(\C{})$ the existence of a compact connected stabiliser 
group $\mathbf K_E \subset U(N)$ can constructively be checked as in Eqn.~\ref{eqn:centraliser}. 
Moreover, every Hermitian $E$ can be chosen diagonal; hence in that case $\mathfrak k_E$ includes 
a maximal torus algebra $\mathfrak t$ with $\mathfrak t \subset \mathfrak k_E \subset\mathfrak{u}(N)$.
The rest follows.
\end{proof}
%%%%%
\begin{remark*}
Obviously, the constraint of leaving $E$ invariant while maximising the
transfer from $A$ to $C$ only makes sense, if $A$ and $E$ do not share
the same stabiliser group.
\end{remark*}

Since it may be tedious to check for the stabiliser group $\mathbf K_E$ of $E$ in each and every
practical instance and then project the gradients onto the corresponding subalgebra $\mathfrak k_E$,
a more versatile programming tool would be welcome.

%%%%%
%%%%%
\begin{algorithm*}
To this type of constrained optimisation, in ref.~\cite{TOSH-Diss},
we also derived a gradient flow based on the \LAG function (with $f_C(U):=\tr\{C^\dagger UAU^\dagger\}$)
\begin{equation}
L(U) = \abs{\TAC}^2 -\lambda\left(\tr\{UE\Ud\Ed\}-||E||_2^2\right)\quad,
\end{equation}
where the constraint $UE\Ud -E = 0$ was written in the more convenient form
        $ \tr\{UE\Ud\Ed\}-\fnormsq{E} = 0$. 
        %let $\ket{\Vec Y}$ and $\ket{\Vec E}$ be of the same norm; then
        %$\braket{\Vec Y}{\Vec E} = \braket{\Vec E}{\Vec E} = \braket{\Vec Y}{\Vec Y}$
        %readily implies the equality of $\ket{\Vec Y}$ and $\ket{\Vec E}$ on
        %geometric grounds. Now set $Y=UE\Ud$.\\
The algorithm implements the gradient from the Fr{\'e}chet derivatives
\begin{equation}
\begin{split}
D\,\big\{\abs{\TAC}^2&-\lambda\TEE+\lambda\,\fnormsq{E}\big\}\,(iHU) \\
        &= \tr\big\{\,\big(2\,(\TACc\GAC )_S - \lambda\GEE\big)\, iH\big\} \quad,
\end{split}
\end{equation}
where for short, $(\cdot)_S$ denotes the skew-Hermitian part,
within the recursion \cite{TOSH-Diss}
\begin{equation}
U_{k+1} = \exp\big\{-\alpha\,\big(2\,(\TACck\GACk )_S - \lambda\GEEk\big)\,\big\} \, U_k\quad.
\end{equation}
\end{algorithm*}
%%%%%%%%%%%%%%%%%%%%%%%%%%%%%%%
%%%%%
%%%%%

\subsubsection{Constraint by Orthogonality}
For the sequel of this paragraph, we assume a generic shape of the $C$"~numerical range \wda so that
$m_0\in\wda$ defines the unique point in \wda that is closest
to the origin in the complex plane. For these instances, we address the optimisation problem
%%%%%
\begin{equation}\label{eq:const_opt}
\maxover U |\tr\{UA\Ud \Cd\}|\quad\text{\rm subject to}\quad \tr\{UA\Ud\Dd\}=m_0 \quad.
\end{equation}
%%%%%
Clearly, perfect matches exist 
if $0\in\wda$, because only then are there points
on the unitary orbit $\mathcal O_u(A)$ that are orthogonal to $D$.
Moreover by Eqn.~\ref{eqn:angle}, the modulus of $m_0$ relates to the cosine of the angle 
between $D$ and points on the unitary orbit $\mathcal O_u(A)$ that come closest to orthogonality.
%Let $m_0\in\wda$ denote such a point closest to the origin in the complex
%plane.
The corresponding $C$-numerical range constrained by (best approximation to) orthogonality to $D$
takes the form \cite{TOSH-Diss}
\begin{equation}
W(C,A)\big|_{\Adr_U \perp D}\,:=\, \big\{\tr\{UA\Ud \Cd\}\,\big|\; |\tr\{UA\Ud \Dd\}| \,%
        =\, m_0 \big\} \subseteq \wca\;,
\end{equation}
which is more difficult to characterise,
because in order to establish whether it is a connected set in the complex plane, 
one has to check the constraint set %of group elements
\begin{equation}
\tilde K_D \,:=\,\{ K \subseteq SU(N)\,|\, \tr\{KAK^\dagger \Dd\}=m_0\;\}
\end{equation}
for its properties of (i) forming a subgroup and (ii) connectedness.
Generically, already the first condition is violated, and only in the rare
event of both of them being fulfilled, the set $\tilde K_D$ turns into a
subgroup ${\mathbf K}_D$,
and hence the constrained $C$"~numerical range
into the relative $C$"~numerical range $W_{{\mathbf K}_D}(C,A)$.
Then the constrained optimisation problem of Eqn.~\ref{eq:const_opt}
would be solved by the relative $C$"~numerical radius
$r_{{\mathbf K}_D}(C,A)$.
Even in the
special case $m_0\,=\,0\in\wda$ %as was the case in Example~\ref{ex:WCAD3} 
it is difficult to make sure the projection of the unitary orbit $\mathcal O_u(A)$
onto the orthocomplement $D^\perp$ of $D$ in the Hilbert space $(\Mat_N(\C{}), \tr\{\cdot^\dagger\cdot\})$ is
still a smooth manifold. Generically, again this is not the case.

It is for these reasons that addressing orthogonality problems by a \LAG approach is
more promising. For this to make sense, one trivially has to ensure $C$ and $D$ are not scalar multiples
of one another, yet is not necessary that perfect orthogonality in the sense of $0\in\wda$
can actually be achieved.

\begin{algorithm*}
In ref.~\cite{TOSH-Diss}, we devised a \LAG-type gradient-flow algorithm for solving
the constrained optimisation problem of Eqn.~\ref{eq:const_opt} numerically.
To this end,
define $\TAC:= \tr\{UA\Ud \Cd\}$ and $\TAD:= \tr\{UA\Ud \Dd\}$.
% where $U:=e^{-itH}$.
Introducing the \LAG function
\begin{equation}
L(U):= \abs{\TAC}^2-\lambda\,\abs{\TAD}^2\quad,
\end{equation}
the task to maximise the transfer from $A$ to $C$ while suppressing the
transfer from $A$ to $D$ can be addressed by implementing the gradient of
\begin{equation}
\begin{split}
D\big\{\abs{\TAC}^2-\lambda\,\abs{\TAD}^2\big\}\,(iHU) &\,=\\ 
 \tr\,\big\{2\, (\TACc\GAC)_S\; iH\big\} &\,- \,\lambda\,\tr\,\big\{2\, (\TADc\GAD)_S\; iH\big\} 
\end{split}
\end{equation}
into the recursive scheme \cite{TOSH-Diss}
\begin{equation}
U_{k+1} = \exp\{-2\alpha\,\big((\TACck\GACk)_S\, -\,\lambda\, \big(\TADck\GADk\big)_S\big)\} \, U_k\quad.
\end{equation}
\end{algorithm*}

%%%%%%%%%%%%%%%%%%%
\begin{figure}[Ht!]
\begin{center}
\includegraphics[width=7cm]{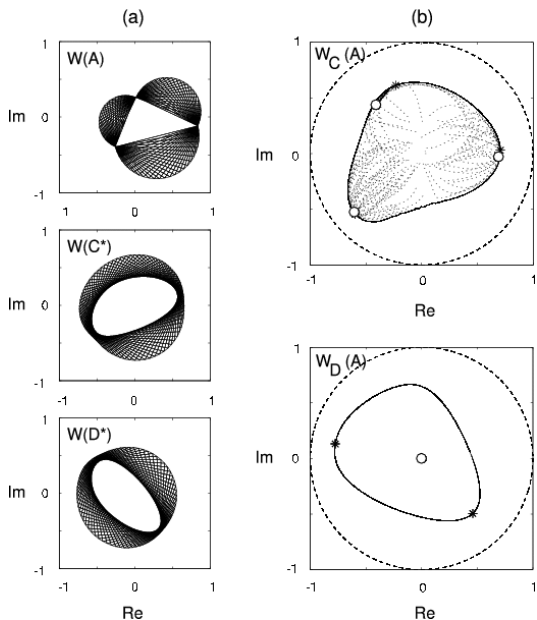}
\end{center}
\caption{\label{fig:L_CAD_3} 
Constrained optimisation $A\to C$ suppressing $A\to D$ for matrices $A,C,D$ of Ex.~\ref{ex:WCAD3}.  
}
\end{figure}
%%%%%%%%%%%%%%%%%%%

\begin{example}\label{ex:WCAD3}
In Fig. \ref{fig:L_CAD_3}, a first example is given for the matrices
\begin{eqnarray}
A &=& {\left(\begin{smallmatrix} %
\pmi 0.8359-0.1152i & 0 & 0\\ %
0 & -0.2593-0.3906i & 0\\ %
0 & 0 & \pmi0.0151+0.2609i %
\end{smallmatrix}\right)}\\[2mm]
C &=& {\left(\begin{smallmatrix} %
   %-0.0318 - 0.0690i     &-0.0404 - 0.0656i &\pmi 0.3086 - 0.1076i \\%
   %-0.3522 + 0.3185i &\pmi 0.0844 + 0.2880i &\pmi 0.1742 + 0.2291i \\%
   %\pmi0.2351 + 0.3050i  &\pmi0.2135 - 0.3234i     &-0.2368 - 0.3585i \\&&\\%
        %%%% 
   -0.0318 + 0.0690i &-0.3522 - 0.3185i & \pmi0.2351 - 0.3050i \\ %  
   -0.0404 + 0.0656i &\pmi 0.0844 - 0.2880i &\pmi0.2135 + 0.3234i \\% 
\pmi0.3086 + 0.1076i &\pmi 0.1742 - 0.2291i    &-0.2368 + 0.3585i %
\end{smallmatrix}\right)}\\[2mm]
D &=& {\left(\begin{smallmatrix} %
  %-0.2910 + 0.3480i &\pmi 0.0836 + 0.2790i  &-0.3906 + 0.1387i \\%
  %-0.2395 - 0.0274i &-0.1836 + 0.0203i      &\pmi0.1989 + 0.2725i \\%
  %-0.2428 - 0.0656i &-0.2427 - 0.2396i      &-0.0442 - 0.3871i \\ && \\
        %
  -0.2910 - 0.3480i &-0.2395 + 0.0274i      & -0.2428 + 0.0656i \\ %
\pmi0.0836 - 0.2790i &-0.1836 - 0.0203i     & -0.2427 + 0.2396i \\ %\pmi0.1989 + 0.2725i \\%
  -0.3906 - 0.1387i &\pmi0.1989 - 0.2725i   &-0.0442 + 0.3871i %
\end{smallmatrix}\right)}
\end{eqnarray}
the first one of which is normal thus entailing a triangular pattern in
its numerical range $W(A)$ shown together with $W(\Cd)$ and $W(\Dd)$
in column (a). The conventional numerical ranges 
were calculated using the classical algorithm of Marcus \cite{MARCUS-87,MARCUS-93,GR-97}. By the constraint that the
transfer from $A$ to $D$ be minimal, the local maxima on the boundary \dwca
of the $C$-numerical range of $A$ (asterisks in column (b))
are shifted to the points indicated by noughts. Note
that only the upper left one seems to be  displaced from the boundary \dwca slightly into
the interior of \wca. All the 100 trajectories (shown as dotted lines)
starting from random
initial conditions converge into the same maxima, while the transfer from
$A$ to $D$ gets zero as shown in the $C$"~numerical range 
$W(D,A)$ at the bottom of column (b).
\end{example}

%%%%%%%%%%%%%%%%%%%%%%%%%%%%%%%
\section{Conclusions}
We have shown how $C$"~numerical ranges provide the setting
for many quantum optimisations. Knowing about its structure paves the way
to numerical algorithms, e.g., for plotting its shape.
In the accompanying paper, we introduced the
relative $C$"~numerical range of $A$ as a restriction of the full unitary group $U$
to some compact connected subgroup $\mathbf K \subset U$, in which case it is connected
but not simply connected. Here we illustrated that this is of practical importance
in relevant examples from quantum information: 
For instance, the maximum real part of the local $C$"~numerical range of $A$
(as a special case of the relative $C$"~numerical range) 
directly corresponds to a measure of \/`pure-state entanglement\/'. In other instances, if the
local $C$"~numerical range of a normalised interaction Hamiltonian $H$ is 
$W_{\rm loc}(H,H)=[-1,+1]$, then the interaction is locally reversible.
These cases are fully characterised in Lie algebraic terms
(by root-space elements to non-zero roots fulfilling a linear system of equations)
%of the non-zero roots of the Lie algebra $\mathfrak{su}(2^n)$
and they are related to cases, where the local $C$"~numerical range is a
circular disc in the complex plane.

Moreover, some constrained optimisation problems in quantum control can be treated
group theoretically: if the constraining conditions can be translated into restricting
the full quantum dynamics on $U$ to a compact connected subgroup $\mathbf K$, 
then the optimisation problem remains within the corresponding connected relative $C$"~numerical range
and the optimisation amounts to finding its relative $C$"~numerical radius.
For more general cases we provide numerical algorithms for constrained optimisations of
\LAG-type.
%%%%%
%%%%%%%%%%%%%%%%%%%%%%%%%%%%%%%
\section{Outlook}
Motivated by applications in quantum control, the relative $C$"~numerical range of $A$ 
introduced awaits further mathematical elucidation: for instance, what are the properties of its boundary,
under which conditions is it simply connected, or even star-shaped? When is it convex? 
How can one systematically find compact connected subgroups embracing practical constraints
of quantum optimisation so that they relate to a connected relative $C$"~numerical range?
What happens in generalisations where the subgroups are no longer compact and connected?
Are there simple instances, in which the correponding restricted $C$"~numerical ranges have few
connected components and thus are not \/`hopeless\/' in view of practical optimisation?
Can one prove the conjecture of ref.~\cite{Science98}
that in generic cases, the gradient flows of Section~\ref{sec:alg_WCA}
always converge to points on the boundary \dwca and there are no local maxima in the interior of \wca? \\
We anticipate that problems in quantum control will profit from mathematical results 
addressing those questions, and---in turn---studying quantum dynamics will stimulate 
conceiving new structures worthy of mathematical research.
%%%%%
%%%%%%%%%%%%%%%%%%%%%%%%%%%%%%%
\vspace{-6mm}
\section*{Acknowledgements}
\vspace{-2mm}
This work was supported in part by the integrated EU-programme QAP.
T.S.H. thanks Chi-Kwong Li and Leiba Rodman for their kind hospitality during
a visit to the College of William and Mary at Williamsburg.

%%%%%%%%%%%%%%%%%%%%%%%%%%%%%%%%%%%%
%\bibliography{control21}

\begin{thebibliography}{10}

\bibitem{LAMA_WKCA}
G.~Dirr, U.~Helmke, M.~Kleinsteuber, and T.~Schulte-Herbr{\"u}ggen.
\newblock { Relative $C$"~Numerical Ranges for Application in Quantum Control and Quantum Information}.
\newblock Accompanying paper in WONRA Proceedings.
\newblock E"~print: http://arxiv.org/pdf/math-ph/0702005, 2007.

\bibitem{DowMil03}
J.P. Dowling and G.~Milburn.
\newblock {Quantum Technology: The Second Quantum Revolution}.
\newblock {\em Phil. Trans. R. Soc. Lond. A}, 361:1655--1674, 2003.

\bibitem{LiTsi91}
 C.K. Li, and N.K.~Tsing.
\newblock {On the $k^{th}$ Matrix Numerical Range}.
\newblock {\em Lin.~Multilin.~Alg.}, 28:229--239, 1991.

\bibitem{Choi++06}
M.D.~Choi, J.A.~Holbrook, D.W.~Kribs, and K.~{\.Z}yczkowsi.
\newblock {Higher-Rank Numerical Ranges of Unitary and Normal Matrices}.
\newblock E"~print: http://arxiv.org/pdf/quant-ph/0608244, 2006.

\bibitem{LiProv93}
N.~Bebiano, C.K. Li, and J.~da~Provid{\^e}ncia.
\newblock {Some Results on the Numerical Range of a Derivation}.
\newblock {\em SIAM J.~Matrix Anal.~Appl.}, 14:1084--1095, 1993.

\bibitem{Science98}
S.~J. Glaser, T.~Schulte-Herbr{\"u}ggen, M.~Sieveking, O.~Schedletzky, N.~C.
  Nielsen, O.~W. S{\o}rensen, and C.~Griesinger.
\newblock {Unitary Control in Quantum Ensembles: Maximising Signal Intensity in
  Coherent Spectroscopy}.
\newblock {\em Science}, 280:421--424, 1998.

\bibitem{TOSH-Diss}
T.~Schulte-Herbr{\"u}ggen.
\newblock {\em Aspects and Prospects of High-Resolution NMR}.
\newblock PhD Thesis, Diss-ETH 12752, Z{\"u}rich, 1998.

\bibitem{NMRJOGO}
U.~Helmke, K.~H{\"u}per, J.~B. Moore, and T.~Schulte-Herbr{\"u}ggen.
\newblock {Gradient Flows Computing the $C$-Numerical Range with Applications
  in NMR Spectroscopy}.
\newblock {\em J. Global Optim.}, 23:283--308, 2002.

\bibitem{DHK06}
G.~Dirr, U.~Helmke, and M.~Kleinsteuber.
\newblock {Lie Algebra Representations, Nilpotent Matrices, and the
  $C$-Numerical Range}.
\newblock {\em Lin.~Alg.~Appl.}, 413:534--566, 2006.

\bibitem{LiWoerd06}
C.K. Li and H.J. Woerdeman.
\newblock {A Lower Bound on the $C$-Numerical Radius of Nilpotent Matrices
  Appearing in Coherent Spectroscopy}.
\newblock {\em SIAM J.~Matrix Anal.~Appl.}, 27:793--800, 2006.

\bibitem{Fey82}
R.~P. Feynman.
\newblock {Simulating Physics with Computers}.
\newblock {\em Int. J. Theo. Phys.}, 21:467--488, 1982.

\bibitem{Fey96}
R.~P. Feynman.
\newblock {\em {Feynman Lectures on Computation}}.
\newblock Perseus Books, Reading, MA., 1996.

\bibitem{Pap95}
C.~H. Papadimitriou.
\newblock {\em Computational Complexity}.
\newblock Addison Wesley, Reading, MA., 1995.

\bibitem{Shor94short}
P.~W. Shor.
\newblock {Algorithms for Quantum Computation}.
\newblock In {\em Proceedings of the Symposium on the Foundations of Computer
  Science, 1994, Los Alamitos, California}, pages 124--134. IEEE Computer
  Society Press, New York, 1994.

\bibitem{Shor97}
P.~W. Shor.
\newblock {Polynomial-Time Algorithms for Prime Factorisation and Discrete
  Logarithm on a Quantum Computer}.
\newblock {\em SIAM J. Comput.}, 26:1484--1509, 1997.

\bibitem{Jozsa88}
R.~Jozsa.
\newblock {Quantum Algorithms and the Fourier Transform}.
\newblock {\em Proc. R. Soc. A.}, 454:323--337, 1998.

\bibitem{Mosca88}
R.~Cleve, A.~Ekert, C.~Macchiavello, and M.~Mosca.
\newblock {Quantum Algorithms Revisited}.
\newblock {\em Proc. R. Soc. A.}, 454:339--354, 1998.

\bibitem{EHK04}
M.~Ettinger, P.~H{\o}yer, and E.~Knill.
\newblock {The Quantum Query Complexity of the Hidden Subgroup Problem is
  Polynomial}.
\newblock {\em Inf. Process. Lett.}, 91:43--48, 2004.

\bibitem{Dav80}
E.B. Davies.
\newblock {\em One-Parameter Semigroups}.
\newblock Academic Press, London, 1980.

\bibitem{GS-77}
M.~Goldberg and E.G. Straus.
\newblock {Elementary Inclusion Relations for Generalized Numerical Ranges}.
\newblock {\em Lin.~Alg.~Appl.}, 18:1--24, 1977.

\bibitem{Li94}
C.-K. Li.
\newblock {$C$-Numerical Ranges and $C$-Numerical Radii}.
\newblock {\em Lin. Multilin. Alg.}, 37:51--82, 1994.

\bibitem{TSING-96}
W.-S. Cheung and N.-K. Tsing.
\newblock {The C-Numerical Range of Matrices is Star-Shaped}.
\newblock {\em Lin.~Multilin.~Alg.}, 41:245--250, 1996.

\bibitem{Li91}
C.-K. Li and N.~K. Tsing.
\newblock {Matrices with Circular Symmetry on Their Unitary Orbits and
  $C$-Numerical Ranges}.
\newblock {\em Proc. Amer. Math. Soc.}, 111:19--28, 1991.

\bibitem{SJ72JS}
H.~Sussmann and V.~Jurdjevic.
\newblock {Controllability of Nonlinear Systems and: Control Systems on Lie
  Groups}.
\newblock {\em J. Diff. Equat.}, 12:95--116 and 313--329, 1972.

\bibitem{GA02}
T.~Schulte-Herbr{\"u}ggen, K.~H{\"u}per, U.~Helmke, and S.~J. Glaser.
\newblock {\em Applications of Geometric Algebra in Computer Science and
  Engineering}, chapter Geometry of Quantum Computing by Hamiltonian Dynamics
  of Spin Ensembles, pages 271--283.
\newblock Birkh{\"a}user, Boston, 2002.

\bibitem{AlbAll02}
F.~Albertini and D.~D'Alessandro.
\newblock {The Lie Algebra Structure and Controllability of Spin Systems}.
\newblock {\em Lin.~Alg.~Appl.}, 350:213--235, 2002.

\bibitem{HJ2}
R.A. Horn and C.R. Johnson.
\newblock {\em Topics in Matrix Analysis}.
\newblock Cambridge University Press, Cambridge, 1991.

\bibitem{Khaneja02}
N.~Khaneja, S.~J. Glaser, and R.~Brockett.
\newblock {Sub-Riemannian Geometry and Time-Optimal Control of Three-Spin
  Systems: Quantum Gates and Coherence Transfer}.
\newblock {\em Phys. Rev. A}, 65:032301, 2002.

\bibitem{PRA05}
T.~Schulte-Herbr{\"u}ggen, A.~K. Sp{\"o}rl, N.~Khaneja, and S.~J. Glaser.
\newblock {Optimal Control-Based Efficient Synthesis of Building Blocks of
  Quantum Algorithms: A Perspective from Network Complexity towards Time
  Complexity}.
\newblock {\em Phys. Rev. A}, 72:042331, 2005.

\bibitem{PRL_decoh}
T.~Schulte-Herbr{\"u}ggen, A.~Sp{\"o}rl, N.~Khaneja, and S.J. Glaser.
\newblock { Optimal Control for Generating Quantum Gates in Open Dissipative
  Systems }.
\newblock E"~print: http://arxiv.org/pdf/quant-ph/0609037, 2006.

\bibitem{PRL_decoh2}
P.~Rebentrost, I.~Serban, T.~Schulte-Herbr{\"u}ggen, and F.K. Wilhelm.
\newblock { Optimal Control of a Qubit Coupled to a Two-Level Fluctuator }.
\newblock E"~print: http://arxiv.org/pdf/quant-ph/0612165, 2006.

\bibitem{NEUM-37}
J.~von Neumann.
\newblock {Some Matrix-Inequalities and Metrization of Matrix-Space}.
\newblock {\em Tomsk Univ. Rev.}, 1:286--300, 1937.
\newblock [reproduced in: {\em John von Neumann: Collected Works}, A.H. Taub,
  Ed., Vol. IV: Continuous Geometry and Other Topics, Pergamon Press, Oxford,
  1962, pp 205-219].

\bibitem{OLE-89}
O.W. S{\o}rensen.
\newblock {Polarization Transfer Experiments in High-Resolution NMR
  Spectroscopy}.
\newblock {\em Prog. NMR Spectroc.}, 21:503--569, 1989.

\bibitem{Bro88+91}
R.~W. Brockett.
\newblock {Dynamical Systems that Sort Lists, Diagonalise Matrices, and Solve
  Linear Programming Problems}.
\newblock In {\em Proc. IEEE Decision Control, 1988, Austin, Texas}, pages
  779--803, 1988.
\newblock see also: Lin. Alg. Appl., 146 (1991), 79--91.

\bibitem{Helmke94}
U.~Helmke and J.~B. Moore.
\newblock {\em Optimisation and Dynamical Systems}.
\newblock Springer, Berlin, 1994.

\bibitem{Guehne04}
J.~Eisert, P.~Hyllus, O.~G{\"u}hne, and M.~Curty.
\newblock {Complete Hierarchies of Efficient Approximations to Problems in
  Entanglement Theory}.
\newblock {\em Phys. Rev. A}, 70:062317, 2004.

\bibitem{Goldbart03}
T.C. Wei and P.M. Goldbart.
\newblock {Geometric Measure of Entanglement and Applications to Bipartite and
  Multipartite Quantum States}.
\newblock {\em Phys. Rev. A}, 68:022307, 2003.

%\bibitem{Lew03}
%K.~Eckert, O.~G{\"u}hne, F.~Hulpke, P.~Hyllus, J.~Korbicz, J.~Mompart,
%  D.~Bruss, M.~Lewenstein, and A.~Sanpera.
%\newblock {\em Quantum Information Processing}, chapter Entanglement Properties
%  of Composite Quantum Systems, pages 79--95.
%\newblock Wiley-VCH, Weinheim, 2003.

\bibitem{BenZyc06}
  I.~Bengtsson and K.~{\.Z}yczkowski.
\newblock {\em Geometry of Quantum States}, 
\newblock Cambridge University Press, Cambridge, UK, 2006.


\bibitem{BruLeu07}
  D.~Bruss and G.~Leuchs, Eds.
\newblock {\em Lectures on Quantum Information}, Section III: Theory of Entanglement,
\newblock Wiley-VCH, Weinheim, 2007.


\bibitem{NC00}
M.~A. Nielsen and I.~L. Chuang.
\newblock {\em Quantum Computation and Quantum Information}.
\newblock Cambridge University Press, Cambridge (UK), 2000.

\bibitem{Hahn50}
E.~Hahn.
\newblock {Spin Echoes}.
\newblock {\em Phys. Rev.}, 80:580--601, 1950.

\bibitem{EBW87}
R.~R. Ernst, G.~Bodenhausen, and A.~Wokaun.
\newblock {\em {Principles of Nuclear Magnetic Resonance in One and Two
  Dimensions}}.
\newblock Clarendon Press, Oxford, 1987.

\bibitem{PRA_inv}
T.~Schulte-Herbr{\"u}ggen and A.~Sp{\"o}rl.
\newblock {Which Quantum Evolutions can be Reversed by Local Unitary
  Operations? Algebraic Classification and Gradient-Flow Based Numerical Checks
  }.
\newblock E"~print: http://arxiv.org/pdf/quant-ph/0610061, 2006.

\bibitem{MARCUS-87}
M.~Marcus.
\newblock {Computer Generated Numerical Ranges and Some Resulting Theorems}.
\newblock {\em Lin.~Alg.~Appl.}, 20:121--157, 1987.

\bibitem{MARCUS-93}
M.~Marcus.
\newblock {\em {Matrices and Matlab: A Tutorial}}.
\newblock Prentice Hall, Englewood Cliffs, N.J., 1993.

\bibitem{GR-97}
K.E. Gustafson and D.K.M. Rao.
\newblock {\em {Numerical Range: The Field of Values of Linear Operators and
  Matrices}}.
\newblock Springer, New York, 1997.

\end{thebibliography}
%%%%%%%%%%%%%%%%%%%%%%%%%%%%%%%%%%%%
%%%%%%%%%%%%%%%%%%%%%%%%%%%%%%%%%%%%

%%%%%%%%%%%%%%%%%%%%%%%%%%%%%%%%%%%%
%%%%%%%%%%%%%%%%%%%%%%%%%%%%%%%%%%%%
%%%%%%%%%%%%%%%%%%%%%%%%%%%%%%%%%%%%
%%%%%%%%%%%%%%%%%%%%%%%%%%%%%%%%%%%%

\label{lastpage}

\end{document}